\documentclass[twocolumn]{aa}
\usepackage{txfonts}
\usepackage{graphicx}
\usepackage{color}
\usepackage{longtable}
\usepackage{natbib}
\bibpunct{(}{)}{;}{a}{}{,} % to follow the A\&A style

\begin{document}

\title{The HARPS search for southern extra-solar planets}
\subtitle{XXXIV. Occurrence, mass distribution and orbital properties of super-Earths and Neptune-mass planets\thanks{Based on observations made with the HARPS instrument on ESO's 3.6\,m telescope at the La Silla Observatory in the frame of the HARPS-GTO Program ID 072.C-0488, the large program for the search of planets around solar-type stars ID 183.C-0972 and the HARPS-Upgrade GTO program ID 69.A-0123}}

\author{M.~Mayor\inst{1}
	\and M.~Marmier\inst{1}
	\and C.~Lovis\inst{1}
	\and S.~Udry\inst{1}
	\and D.~S\'egransan\inst{1}
	\and F.~Pepe\inst{1}
	\and W.~Benz\inst{2}
         \and J.-L.~Bertaux\inst{3}
	\and F.~Bouchy\inst{4}
	\and X.~Dumusque\inst{1}
	\and G.~LoCurto\inst{5}
	\and C.~Mordasini\inst{6}	
	\and D.~Queloz\inst{1}
	\and N.C.~Santos\inst{7,8}
	}
	
\offprints{M. Mayor, \email{Michel.Mayor@unige.ch}}
	   
\institute{Observatoire Astronomique de l'Universit\'e de Gen\`eve, 51 ch. des Maillettes - Sauverny, CH-1290 Versoix, Switzerland
          \and
          Physikalisches Institut, Universit\"at Bern, Silderstrasse 5, CH-3012 Bern, Switzerland
          \and 
          LATMOS, CNRS/UVSQ Universit\'e de Versailles, Saint Quentin, France
          \and
          Observatoire de Haute-Provence, 04870 Saint-Michel l'Observatoire, France
          \and
          European Southern Observatory, Karl-Schwarzschild-Str. 2, D-85748 Garching bei M\"unchen, Germany            
          \and 
          Max-Planck-Institut f\"ur Astronomie, K\"onigsstuhl 17, D-69117 Heidelberg, Germany  
          \and
          Centro de Astrof\'{\i}sica , Universidade do Porto, Rua das Estrelas, 4150-762 Porto, Portugal
           \and
	 Departamento de F\'isica e Astronomia, Faculdade de Ci\^encias, Universidade do Porto, Portugal
	 }

\date{Received  September 2011 / Accepted 2011}

\abstract
   {}
   {We report on the results of an 8-year survey carried out at the La Silla Observatory with the HARPS spectrograph to detect and characterize planets in the super-Earth and Neptune mass regime. }
   {The size of our star sample and the precision achieved with HARPS have allowed the detection of a sufficiently large number of low-mass planets to study the statistical properties of their orbital elements, the correlation of the host-star metallicity with the planet masses, as well as the occurrence rate of planetary systems around solar-type stars. }
   {A robust estimate of the frequency of systems shows that more than 50\% of solar-type stars harbor at least one planet of any mass and with period up to 100\,days. Different properties are observed for the population of planets less massive than about 30\,M$_{\oplus}$ compared to the population of gaseous giant planets. The mass distribution of Super-Earths and Neptune-mass planets (SEN) is strongly increasing between 30 and 15\,M$_{\oplus}$. The SEN occurence rate does not exhibit a preference for metal rich stars. Most of the SEN planets belong to multi-planetary systems. The orbital eccentricities of the SEN planets seems limited to 0.45. At the opposite, the occurence rate of gaseous giant planets is growing with the logarithm of the period, and is strongly increasing with the host-star metallicity. About 14\,\% of solar-type stars have a planetary companion more massive than 50\,M$_{\oplus}$ on an orbit with a period shorter than 10~years. Orbital eccentricities of giant planets are observed up to 0.9 and beyond. } 
   {The precision of HARPS-type spectrographs opens the possibility to detect planets in the habitable zone of solar-type stars. Identification of a significant number of super-Earths orbiting solar-type of the Sun vicinity is achieved by Doppler spectroscopy. 41 newly discovered planets with HARPS are announced in the Appendix of this paper, among which 16 Super-Earths.}

\keywords{Stars: late-type -- Planetary systems -- Techniques: radial velocities -- Techniques: spectroscopy -- Methods: statistical analysis}

\maketitle
%________________________________________________________________

%----------------------------------------------------------------
\section{Super-Earths and Neptune-mass planets around solar-type stars}
%----------------------------------------------------------------

Thanks to the increasing sensitivity of instruments optimized for the measurement of stellar radial velocities, planets with masses smaller than 2\,M$_\oplus$ have been detected. Over the past few years an impressive progress has been made for the detection of close-in super-Earths (planets with masses between 1 and 10\,M$_\oplus$) and Neptune-mass planets. At present, the number of low-mass planets discovered is sufficient to study the statistical properties of this rich sub-population and to estimate their occurrence frequency around solar-type stars.
 
The first hint to low-mass planets orbiting solar-type stars on tight orbits was provided by the discoveries of the short period companions hosted by 55\,Cnc \citep{McArthur:2004} and $\mu$\,Ara \citep{Santos:2004b}. Both discoveries were the result of intensive and  special-interest Doppler monitoring: the study of the complex dynamics for the first system and an asteroseismology campaign for the second. Orbital solutions and masses for these two low-mass planets have been subsequently revised as more measurements were gathered and new approaches applied for the data analysis. For 55\,Cnc\,e, \citet{Dawson:2010} analyzed the impact of observation aliases on the solution determination and proposed a revised value for the period passing from 2.7 down to $P=0.736$\,days corresponding to a new minimum mass $m_2\sin{i}=8.3$\,M$_{\oplus}$) for the planet. In the case of $\mu$\,Ara \citet{Pepe:2007} revised the mass of the planet at $P= 9.6$\,days to $m_2\sin{i}=10.5$\,M$_{\oplus}$ in the context of a new orbital solution implying a total of four planets. These first detections induced important changes in our observing strategy, leading now to the discovery of a wealth of low-mass planets with short orbital periods. Measurements were made long enough to diminish the stellar acoustic noise, as well as series of measurements over consecutive nights to lower granulation noise, were the key for demonstrating the existence of an extremely rich population of low-mass planets on tight orbits, these planets being frequently members of multi-planetary systems. In this context, a few landmark contributions of HARPS to the field can be recalled:
\begin{itemize}
\item[--] In 2006, the discovery of three Neptune-mass planets hosted by HD\,69830 \citep{Lovis:2006}, an interesting K star surrounded by a prominent disk of dust.
\item[--] In 2009, the discovery of a planetary system with three super Earths orbiting HD\,40307 \citep{Mayor:2009a}. 
\item[--] In 2011, the characterization of a system with 7 planets on tightly-packed orbits around HD\,10180 Ê\citep{Lovis:2011a}. This is the planetary system with the largest number of planets known so far, with the exception of our own Solar System.
\item[--] After 4 years of the HARPS survey \citep{Mayor:2003}, the number of low-mass planets detected was large enough to allow for a preliminary analysis of the statistical properties of this specific population \citep{Udry:2007a,Mayor:2008}. The mass distribution of small-mass planets was found to be rising towards smaller masses down to 10\,M$_{\oplus}$, where the measurements start to be dominated by detection limits. A bi-modal mass distribution is observed for low-mass planets and giants. Concerning the planet-star separation, the pile-up around 3 days observed for hot Jupiters was not observed for the low-mass population \citep{Udry:2007a}. Already with the very first detections of low-mass planets, it appeared that these planets were hosted by stars without overabundance of metals \citep{Udry:2006}, in heavy contrast with the strong correlation between host-star metallicity and occurrence frequency observed for gaseous giants \citep{Santos:2001,Santos:2004a,Fischer:2005}. This specific point was revisited and strengthened by \citet{Sousa:2008} who carried out a homogenous metallicity determination of the HARPS sample stars.
\item[--] A first estimate of the occurrence rate of low-mass planets (e.g. below 30\,M$_{\oplus}$, with periods shorter than 50 days) was proposed by \citet{Lovis:2009}. The estimated frequency was $\sim30$\,\%. Note that, based on the $\eta$\,Earth project carried out at Keck, a significantly lower estimate for the occurrence rate of low-mass planets was proposed by \citet{Howard:2010}. Their estimate was of $\sim18$\,\% for planets between 3 and 30\,M$_{\oplus}$ and periods shorter than 50 days. Their analysis included a correction for detection biases.
\end{itemize}
\noindent

Recently, the Kepler space mission contributed to significantly to our knowledge of low-mass planet properties by publishing a list of 1235 transit detections of small-size planet candidates \citep{Borucki:2011}. In addition to the confirmation of the existence of the rich population of low-mass planets on tight orbits, a fantastic result of the mission resides in the high frequency of multi-transiting systems in extremely co-planar configurations. As for the HARPS survey, the deduce that a large number of stars host several planets. A stunning example of such systems is given by the 6 transiting planets around Kepler-11 \citep{Lissauer:2011}.

The present study mainly discusses the properties of the population of low-mass planets hosted by solar-type stars. We note, however, that this sub-population also exists around M stars, as illustrated for example by the four-planet system around GJ\,581\footnote{\citet{Vogt:2010} announced the detection of 2 additional planets in the Gl\,581 system, with periods of 36.6 and 433 days, based on the analysis of the 119 published HARPS radial velocities \citep{Mayor:2009b} and 122 Keck HIRES measurements. With a final set of 240 high-precision HARPS measurements, \citet{Forveille:2011} are now able to rule out the existence of the claimed 2 additional planets in the system.} \citep{Bonfils:2005,Udry:2007b,Mayor:2009b,Forveille:2011}. For a discussion of the HARPS companion subprogram devoted to M stars we refer to \citet{Bonfils:2011}.

The HARPS and CORALIE surveys used in the present study are described in Sect.\,2, with a discussion of the stellar samples, the corresponding radial-velocity precision, and the correction of the detection biases. The planetary systems at the basis of our discussion will be presented in Sect.\,3 and the derived distributions of planetary masses, periods, and eccentricities in Sect.\,4.  The emphasis of this study is on the results obtained for the population of super-Earths and Neptune-mass planets. However, combining our HARPS data with results of the CORALIE planet-search program allows us to derive statistical distributions for the whole domain of planetary masses, from less than two Earth masses up to the most massive giant planets. This approach will in particular allow us to compare the derived statistical properties as a function of planetary mass. Clearly, with the radial-velocity technique, the detection of low-mass planets on longer periods is strongly affected by detection biases. These detection biases have been carefully modeled and corrected. This bias correction is of great importance when we estimate the occurrence of planets as a function of mass and period. This two-variable distribution is one of the most direct way to compare observations and Monte-Carlo based planet population synthesis as e.g. developed by \citet{Ida:2004,Ida:2005,Ida:2008a,Ida:2008b} or \citet{Mordasini:2009a,Mordasini:2009b}. The physics of planetary formation dependents on the metallicity of the accretion disk. The large number of super-Earths and Neptune-mass planets discovered in our survey provides a unique sample to derive the distribution of the host-star metallicity as a function of planet masses (Sect.\,5). One of the most important goals of the field is certainly the estimation of the number of Earth-type planets lying in the habitable zone of solar-type stars. Still more important would be the preparation of a list of stars with such planets at short distance to the Sun, preparing for follow-up studies aiming at detecting life signatures on those planets. High-precision Doppler surveys will contribute to the build up of such an "input catalogue" (see Sect.\,6).

%----------------------------------------------------------------
\section{The HARPS and CORALIE planet-search programmes}
%----------------------------------------------------------------
\subsection{Stellar samples} 
%----------------------------------------------------------------
\subsubsection{The CORALIE survey}
%----------------------------------------------------------------
A volume-limited radial-velocity survey is going on since 1998 with the CORALIE spectrograph, located at the Nasmyth focus of the 1.2-m EULER Swiss telescope at La the Silla Observatory. CORALIE is a twin of the ELODIE spectrograph \citep{Baranne:1996} running in the past at the Haute-Provence Observatory, and having allowed the discovery of 51\,Peg\,b \citep{Mayor:1995}. In spite of the small telescope size, this instrument has a significant share of exoplanet discoveries. The volume-limited survey includes some 1650 stars of the southern sky. The limit distance for the stars in the sample is depending on the stellar spectral type: it has been set at 50\,parsecs for the F and G dwarfs, and linearly decreasing between K0 and M0, in order to keep a reasonable exposure time to reach a given precision \citep{Udry:2000}. During the first ten years of observations, between 1998 and 2008, the typical precision of CORALIE radial-velocity measurements was about 5\,ms$^{-1}$. In 2008, an upgrade of the spectrograph allowed for an improvement of its efficiency and precision. The goal of the CORALIE survey is to get, for an unbiased volume-limited sample, properties of substellar companions of solar-type stars, from sub-Saturn mass giant planets up to the binary star regime. Some interesting results have already been obtained from this material as e.g. the correlation between host star-metallicity and occurrence frequency of giant planets \citep{Santos:2001,Santos:2004a}, or a better definition of the location of the brown-dwarf desert, and hence of the observed maximum mass of giant planetary companions at intermediated periods \citep{Sahlmann:2010}.

%----------------------------------------------------------------
\subsubsection{The HARPS survey}
%----------------------------------------------------------------
The HARPS instrument is a high-resolution spectrograph optimized for precise radial-velocity measurements, and specially developed for exoplanet search programs \citep{Mayor:2003}. Temperature and pressure are controlled in a vacuum tank located in a thermal enclosure, the instrument is located in the gravity invariant Coud\'e room and is fed with optical fibers issued from the Cassegrain adapter at the telescope. Thanks to its extreme thermal stability and the stable illumination of the optics provided by optical fibers (including a double-scrambling device), an instrumental precision better than  1\,ms$^{-1}$ is routinely achieved \citep{Pepe:2008}. Both instruments (CORALIE and HARPS) are determining radial velocities by using the so-called cross-correlation technique \citep{Baranne:1996} between the extracted and calibrated high-resolution spectra and a binary (0,1) template. The combination of the instrument and the technique is very efficient. For illustration, HARPS  on the 3.6-m telescope reaches, on a 7.5 magnitude star, a precision of 1\,ms$^{-1}$  in only one minute. The dispersion relation for the orders of the stellar spectra on the CCD are determined by using a thorium-argon calibration lamp. A second parallel optical fiber is used for a simultaneous measurement of the calibration lamp during science exposures, monitoring thus potential drifts of the spectrograph during the night, and allowing subsequently for a correction of this drift if necessary. However, the nightly drift is typically much smaller than 1\,ms$^{-1}$ in HARPS.

As a return for the construction of the instrument, 500~nights of guaranteed-time observations (GTO) have been granted by ESO to the HARPS scientific consortium in order to conduct a comprehensive search for exoplanets in the southern sky. As part of the different sub-programs carried out within the GTO since October 2003, a fraction of 50\,\% of the GTO time has been dedicated to the exploration of the domain of very-low mass planets around solar-type stars (from late-F to late-K dwarfs), a domain of masses which did not yet present any detections in 2003. This program proved to be extremely successful and, after the end of the allocated GTO observing time (PI: M.Mayor) in April 2009, it has been continued as an ESO Large Program (PI: S. Udry). The targets followed at high-precision with HARPS have been selected as a subsample of the CORALIE volume-limited sample, with additional constraints however: F and G stars with chromospheric activity index larger than $\log{R'_{HK}}=-4.75$ (or larger than $-4.70$ for K stars) or high-rotation star have been rejected, as well as binaries. We were finally left with a sample of 376 supposedly non-active stars. 

An additional exploratory program aiming at finding still smaller-mass planets around a small sub-sample of 10 stars in the GTO high-precision survey by applying a more demanding observing strategy to average the perturbing effects of stellar intrinsic phenomena (pulsations, granulation, activity) on the radial-velocity measurements \citep{Dumusque:2011a} is also being carried out on HARPS (PI: F.Pepe). This program has already demonstrated the capability of HARPS to detect smaller-amplitude planets \citep{Pepe:2011} as, e.g., a planet with radial-velocity amplitude as small as 0.56\,ms$^{-1}$ and a planet in the habitable zone of a K dwarf (see also Sect.\,6). 

The statistical results presented in the following are derived from the combined available data from these programs as per July 2011. Altogether the analyzed data are based on more than 450 observing nights with the HARPS spectrograph .

%----------------------------------------------------------------
\subsubsection{A volume-limited  sample for the analysis}
\begin{figure}
  \resizebox{\hsize}{!}{\includegraphics{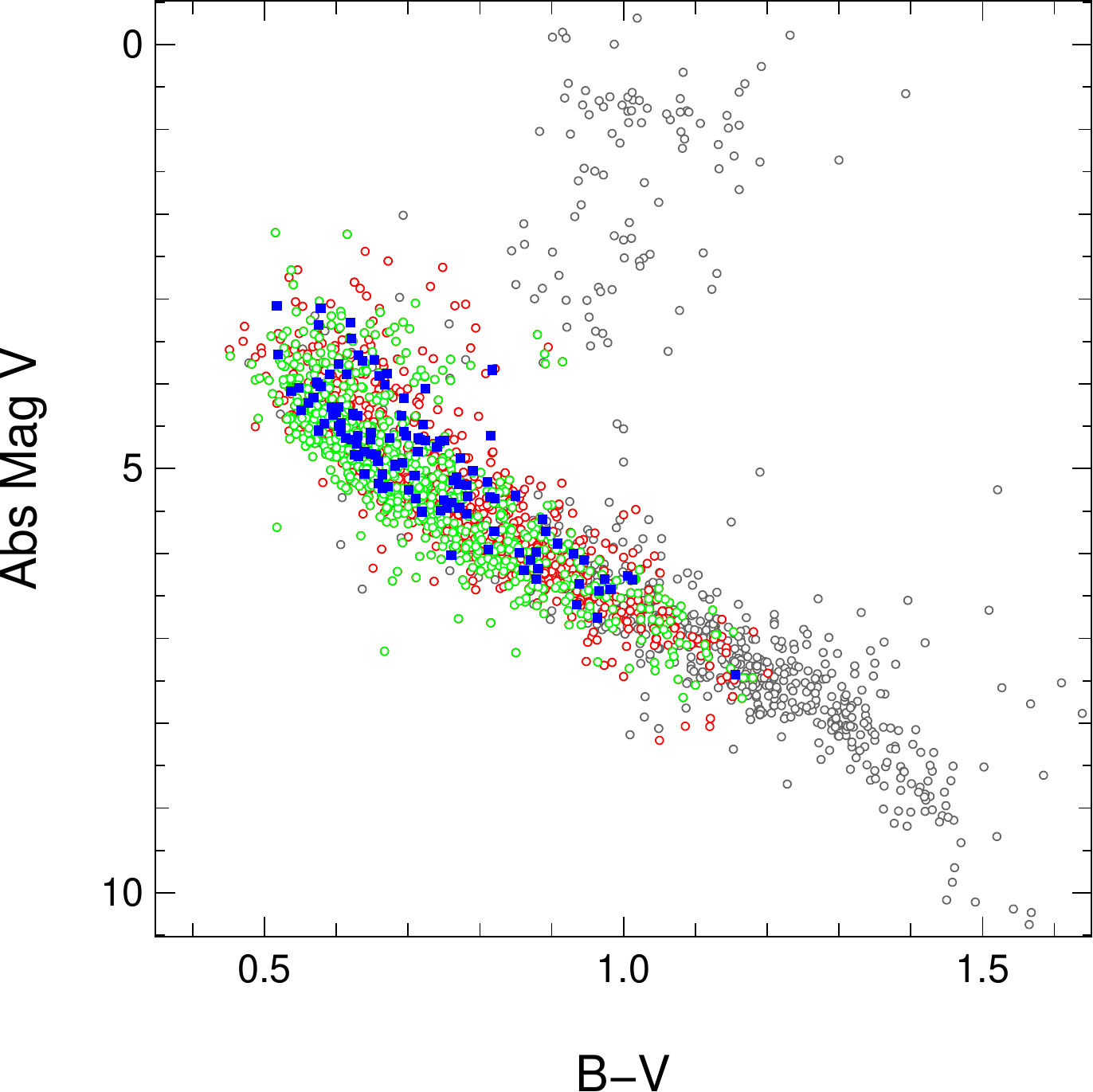}}
  \caption{HR-diagram of Hipparcos 2008 catalogue (black) , CORALIE volume limited sample (red), HARPS  sample of low activity stars (green) and stars with planetary systems (blue)}
  \label{HARPS_MM_HRdiag}
\end{figure}

%----------------------------------------------------------------
As the data are delivered by two (or three, if we consider the Coralie Upgrade as a new instrument) different instruments with measurements of different precisions, as well as of different observation time spans and strategies, we adopted a statistical treatment taking into account these various constraints and allowing us to use in an optimum way the information content of each of the surveys. The CORALIE survey, with its more modest radial-velocity precision but longer duration, mostly provides information about gaseous giant planets with orbital periods covering a long time interval from below 1~day to more than 12~years. The HARPS survey, on the other hand, provides a very deep insight into the domain of Super-Earths and Neptune-type planets up to periods of about 100 days. To avoid as much as possible differential biases between the 2 samples, the CORALIE sample used in the analysis has been re-defined following the same criteria for non-active star selection as for the HARPS sample. As stellar activity does not a priori correlates with distance to the Sun, the sample keeps its statistically well-defined characteristic. Finally, the global sample for the statistical analysis includes 822 non-active stars allowing to estimate detection limits, observational biases, and correct for them in a large domain in the $m_2\sin{i} - \log{P}$ plane. The HR-diagram of the final sample is shown in Fig.\,\ref{HARPS_MM_HRdiag}.

%----------------------------------------------------------------
\subsubsection{Detection limits}
\label{sect_detlimit}
%----------------------------------------------------------------
Since every star in the statistical sample has a different observing history, we computed, as a function of period, the maximum planet mass of a potential planet which still compatible with the RV measurements of this same star. For \emph{each star} (with or without a detected planet candidate) the so computed detection limit determines if it usefully constrains the distribution at a given mass and period. Several authors \citep{Cummings:2007,Zechmeister:2009,Howard:2010,Bonfils:2011} have presented statistical analysis of their planet-search program and discussed methods for finding unbiased distributions consistent with a set of detections. They all assume circular orbits to estimate the upper limits on the amplitude. We also adopt this safer approach to avoid problems in case of inadequate phase coverage. Moreover, \citet{Endl:2002} have shown that the detectability is only slightly affected by this hypothesis for eccentricities below $e=0.5$. The originality of our method is to combine for each star up to three data sets from the HARPS and CORALIE spectrographs\footnote{The efficiency of CORALIE was increased in 2007 through a hardware optimization that has affected the radial velocity zero point. For this reason, we consider the upgraded spectrograph as a new instrument.}. In order to constrain the radial velocity zero point, times series with at least four points on each instrument are combined. If the star is known to host a planetary system, the velocity offset is left as a free parameter of the fitting process. Otherwise a $\chi^{2}$ test is used to determine if a linear trend represents a significant improvement compared to the constant model. After combining the data sets and removing the identified planetary signals, the stellar activity correlation is removed using the relation with the $\log R'_{HK}$ described in \citet{Lovis:2011b}. We obtain this way a single set of residuals per star which has been cleaned from detected planetary signatures, identified activity cycles and instrumental offsets. A search for a residual excess of periodic variability is then carried for every star.

We followed the procedure described in \citet{Zechmeister:2009}: We used the generalized Lomb-Scargle (GLS) periodogram to explore FAP level above $10^{-3}$ over 5000 bootstrap randomization of the data set. Whenever a possible periodicity in the RV data is unveiled at an FAP of 1\%, a careful search for correlation with stellar parameters ($\log R'_{HK}$, $FWHM$, line bisector) is carried on. If the signal doesn't seem to be related to the stellar activity, we use its period as a guess for a Keplerian fit with \textit{Yorbit} \citep{Segransan:2011}. If no credible orbit can be found or if a correlation with stellar jitter is suspected, the signal is removed by fitting a sine wave. Otherwise, the planet orbit is subtracted and the signal flagged as possible planetary candidate to be analyzed further. The whole procedure is re-applied on the so-obtained new residuals until no more significant signal can be found. It must be note that planet candidates found during this procedure (a total of 6) are considered for the statistical study but are non listed in the table of detected planets (see Table\,\ref{mayor_tab1}). 

After this operation we are able to derive for every star the period-mass limit above which we can exclude or detect a planet. As all the significant signals where removed from the time series, the residuals can be considered as pure noise. On these residuals we first apply the GLS on 5000 bootstrap randomizations of the data set to determine a power threshold corresponding to 1\% FAP. The presence of a planet at the trial period $P$ is then mimicked by adding a sine wave to the data. The calendar of measurements remains unaffected. The planet is considered to be detected if the power in the GLS periodogram at the period $P$ (or at its 1 day alias) is greater than the threshold. Using the corresponding sine wave amplitude and the stellar mass, we can deduce the mass corresponding to a 99\% detection limit. Due to the imperfect sampling of the measurements, the detection limit depends in practice on the phase of the signal. Therefore we compute the detection limits over 20 equidistant phases between 0 and 1 and treat them as 20 different stars. This solution provides a simple way of 'averaging' over phases with variable detection limits.

%----------------------------------------------------------------
\subsubsection{Completeness of the survey and planetary rate}
%----------------------------------------------------------------
The global efficiency of the survey, as represented in Fig.\ref{HARPS_MM_m2sini-P_observations} and \ref{HARPS_MM_m2sini-P_3-100M_Pinf1year}, is deduced from the upper mass envelope of the independent limits of the sample. For example, the 20\% line represents the limit below which only 20\% of the stars have a 99\% detection. Such a diagram displays the fraction $C(m\sin i, P)$ of stars with sufficient measurements to detect or exclude a planet at a given period and mass. In other terms it tells us what is the subsample of stars useful to estimate the planetary rate in any mass-period region and how many planetary companions were potentially missed due to insufficient constraints provided by the observations. In order to determine and compare the planetary occurrence rate in different regions of the diagram, we assign an effective number $N_{i,j} = 1/C(msini,P)$ for each detected planet $i$ around the star $j$. The planetary rate is then simply given by $f_{pl} = \frac{1}{N_{*}} \sum_{i}{N_{i,j}}$ where $N_{*}$ is the total number of stars in the sample. The sum is calculated over every detected planet $i$ in a given mass-period domain. Computing the occurrence of planetary systems is similar, using $f_{syst} = \frac{1}{N_{*}} \sum_{j}{N_{max,j}}$, except for the sum which is calculated on every host star $j$ accounting only for the companion with the largest effective number $N_{i,j}$ in the trial domain.

%----------------------------------------------------------------
\section{Planetary systems in the combined sample} 
%----------------------------------------------------------------
155 planets belonging to 102 planetary systems have been identified among the stars in our combined sample. More than two third of these planets have been discovered by the CORALIE and HARPS surveys. HARPS or CORALIE orbits also exist for most of the planets in our sample which were discovered in the frame of other programs. The list of the 102 planetary systems with orbital parameters and references to the discovery papers is provided in Table\,\ref{mayor_tab1}. The characteristics of the newly discovered planets and planetary systems included in the statistical analysis are furthermore presented in the appendix of this paper. The detailed analysis of these systems will be presented however in the papers referred to in Table\,\ref{mayor_tab1}. For the statistical analysis, the 6 additional 'planetary candidates' issued from the computation of the detection limits have been included.

\begin{figure}
  \resizebox{\hsize}{!}{\includegraphics{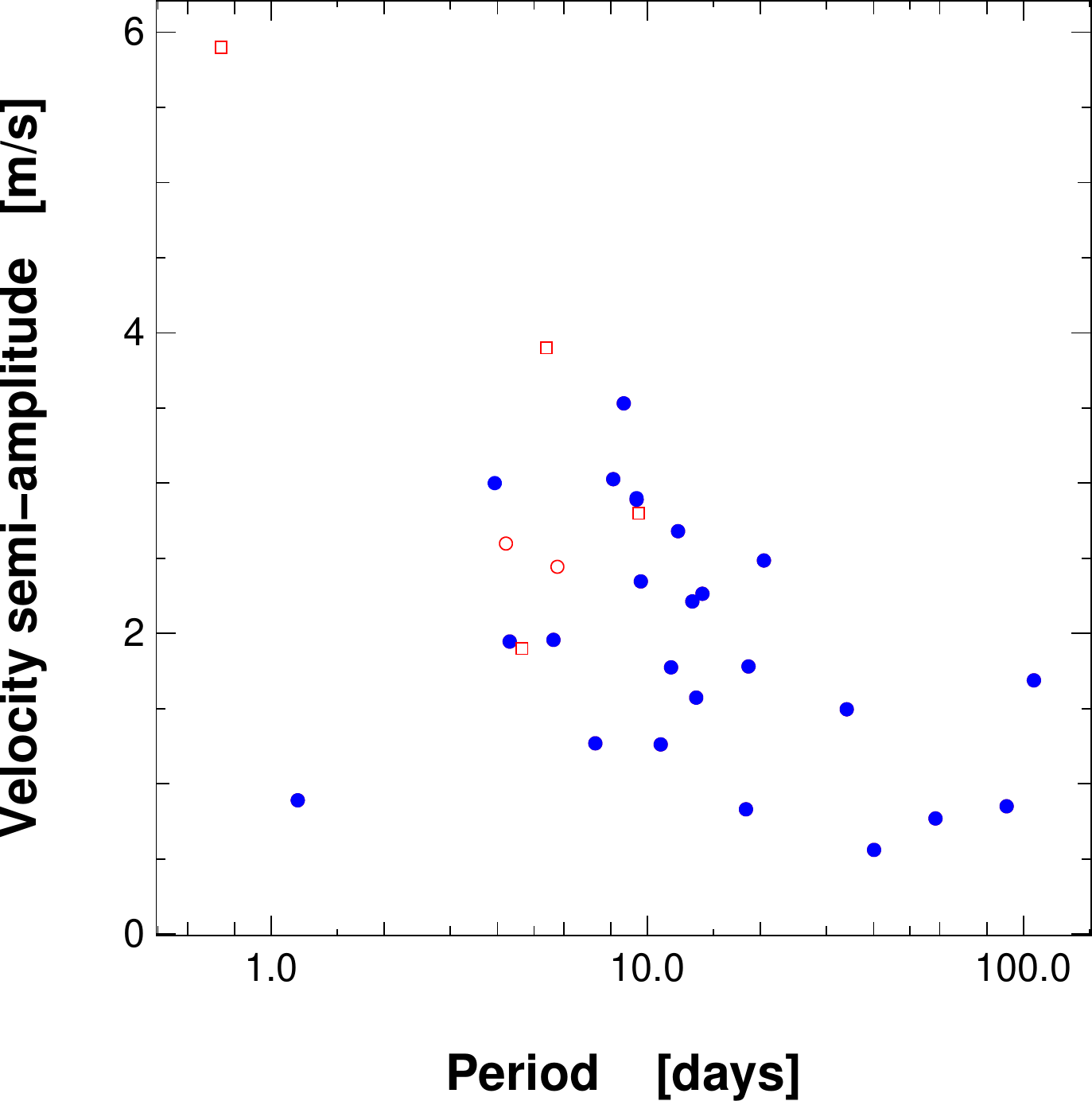}}
  \caption{Radial-velocity semi-amplitude $K$ as a function of orbital period for super-Earths ($M<10$\,M$_{\oplus}$) hosted by solar-type stars. HARPS detection are plotted as blue dots and objects from the literature in red symbols (circles for the southern sky and square for the northern sky).}
  \label{HARPS_MM_K-P}
\end{figure}

A key parameter for the statistical discussion of low-mass planets is the global efficiency of the Doppler measurements in a given survey. This efficiency includes the precision of the spectrograph, sample star characteristics, and the observing strategy (including measurement frequency, exposure time long enough to diminish the different intrinsic stellar noises, and photon noise). In Fig.\,\ref{HARPS_MM_K-P}, we plot the representative points of the radial-velocity semi-amplitudes $K$ as a function of the logarithm of the period $\log{P}$ for all super-Earths orbiting solar-type stars. Planets detected in the frame of other surveys are identified by red symbols and those of our surveys by blue dots. This plot is a straightforward illustration of the contribution of HARPS in this field. The number of the newly detected super-Earths is impressive, as well as the very small amplitudes presently detected by the instrument. This diagram shows how important the measurement precision turns out to be for the correct estimation of the occurrence of low-mass planets, which may be strongly biased if the low-mass planet detections delivered by HARPS were lacking.  

\begin{figure}
  \resizebox{\hsize}{!}{\includegraphics{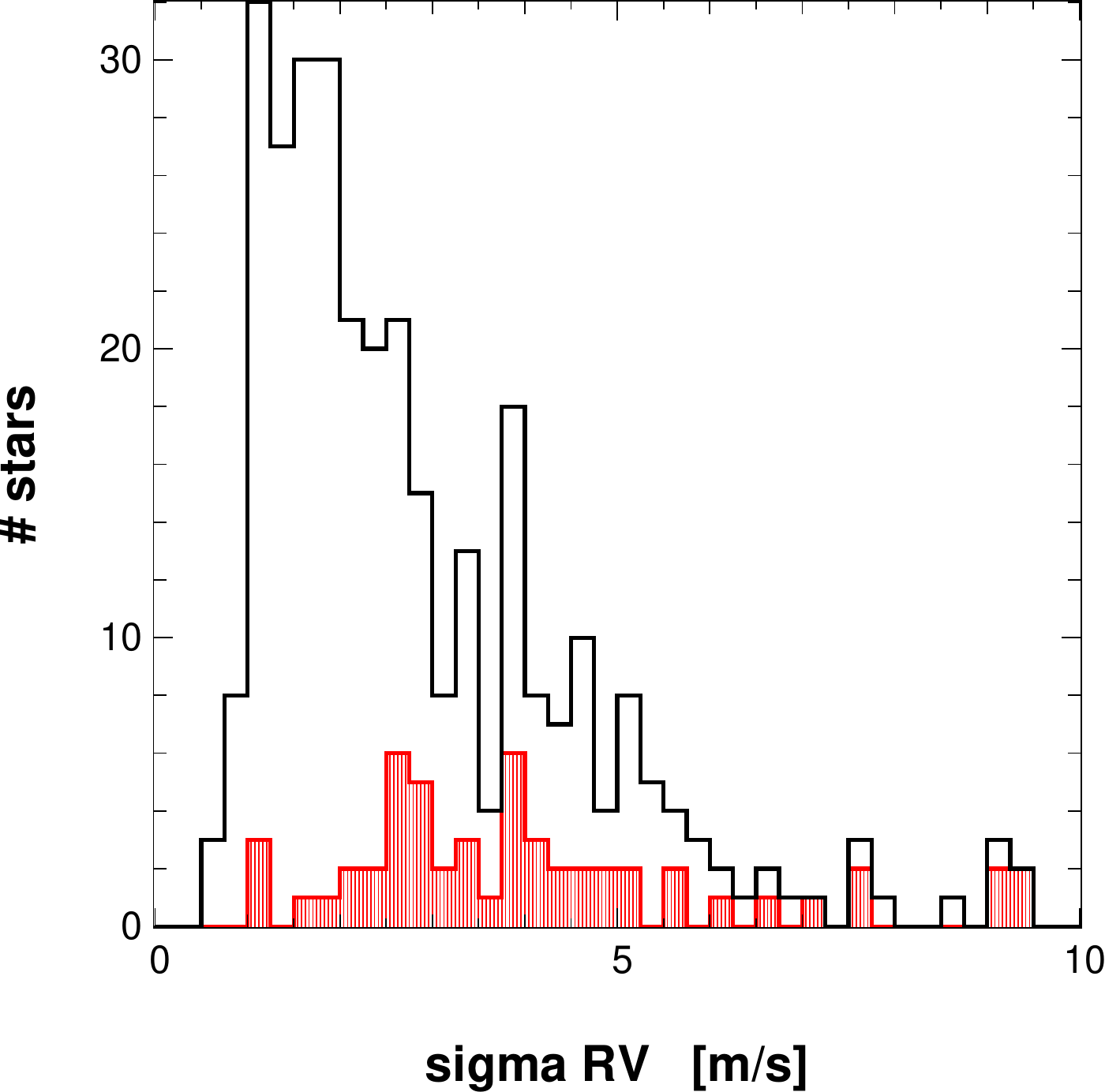}}
  \caption{Histogram of the radial-velocity dispersion for the stars of the HARPS sample (black line). Only the part of the histogram with $rms <10$\,ms$^{-1}$ is displayed. The red histogram represents the \emph{raw} radial-velocity dispersion for stars with detected exoplanets.}
  \label{HARPS_MM_histo_rmsVr}
\end{figure}

\begin{figure}
  \resizebox{\hsize}{!}{\includegraphics{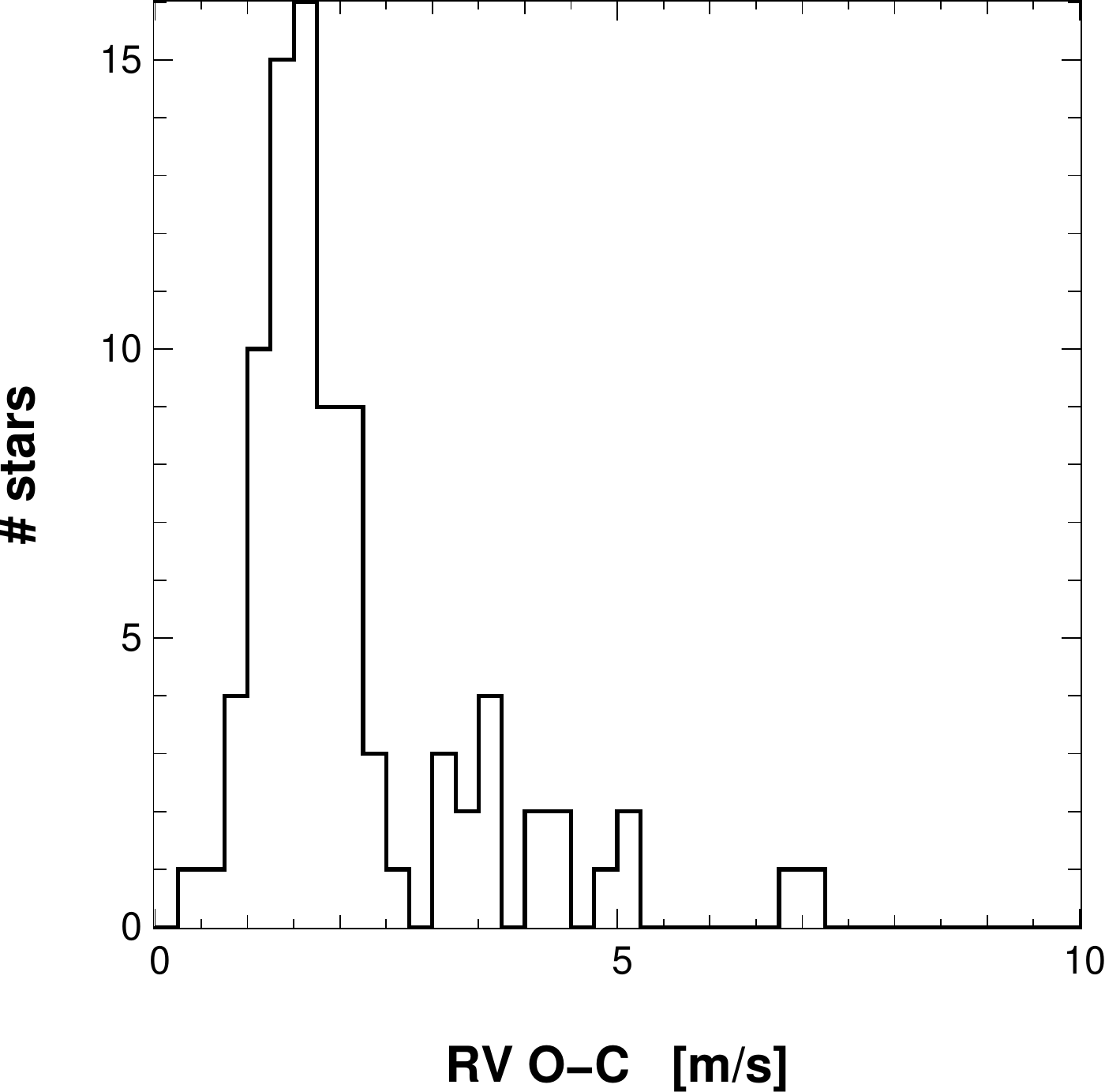}}
  \caption{Histogram of the $O-C$ residuals after fitting the (multi-) Keplerian model to the velocity measurements of a star with planets. The $O-C$ dispersions include all possible sources of noises (photon noise, instrumental errors and stellar jitter), as well as the signal from still undetected planets. Note that a significant fraction of the $O-C$ lies below 1\,ms$^{-1}$.}
  \label{HARPS_MM_histo_O-C}
\end{figure}

Figure\,\ref{HARPS_MM_histo_rmsVr} presents the radial-velocity raw dispersion measured on the HARPS targets (only the histogram of $rms$ less than 10\,ms$^{-1}$ is displayed). Superimposed, we have also illustrated the histogram of the $rms$ for stars with detected planetary systems. We should note that some planetary systems are detected around stars having quite modest \emph{raw} radial-velocity scatter, in some cases smaller than 2\,ms$^{-1}$. Once again, the importance of the measurement precision turns out to be of fundamental importance when exploring the domain of super-Earth masses. After fitting planetary orbit(s) to our set of measurements we derive the residuals around the orbital solutions ($O-C$). These residuals include all potential sources of noise, as instrumental errors, photon noise, stellar intrinsic noise (acoustic oscillation, granulation, activity, magnetic cycle), as well as still undetected small-mass planets. Magnetic cycles, analogues to the Sun's 11-year cycle, can furthermore induce variations of the radial velocities over several years and amplitudes up to 10\,ms$^{-1}$ \citep{Lovis:2011b}. This effect is observed also for some stars with rather low activity level. Conveniently however, several spectral signatures ($\log{R'_{HK}}$, width of the cross-correlation function (CCF) or spectral line and CCF bisectors) can and must be used to identify magnetic cycles and correct for their influences on the velocities \citep{Lovis:2011b, Dumusque:2011b}. These spectral signatures can also be used to test velocity variations on shorter time scales (days or weeks) to detect the possible influence of spots or other anisotropic features on the stellar surface. These diagnostics are of particular importance when searching for low-amplitude velocity variations, typical of low-mass planets. All these tests have been applied to the HARPS and CORALIE  planet-host stars and are described in the corresponding discovery papers. The histogram of the $O-C$ (Fig.\,\ref{HARPS_MM_histo_O-C} for the HARPS planetary systems is certainly the most significant estimation of the global precision of the program.  The mode of the $O-C$ distribution lies at 1.4\,ms$^{-1}$ but many planetary systems show $O-C$ smaller than 1\,ms$^{-1}$ even with a large number of measurements spanning several years of observations.

%----------------------------------------------------------------
\section{Statistical analysis of the sample}
%\section{Distributions of planetary masses, orbital periods  and occurrence of planets}
%----------------------------------------------------------------
\subsection{The mass-period distribution} 
\begin{figure}
  \resizebox{\hsize}{!}{\includegraphics{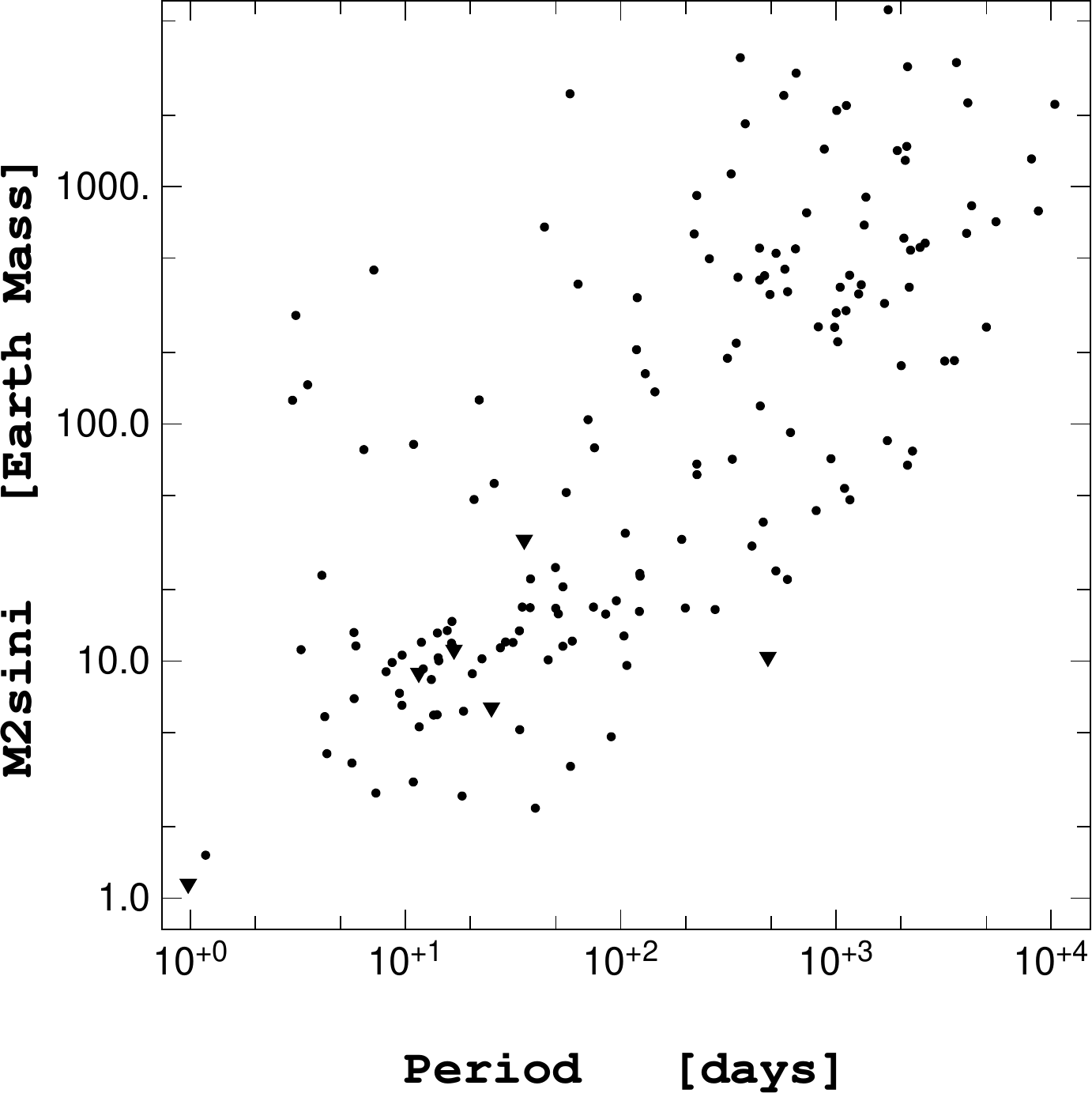}}
  \caption{Plot of the 155 planets (dots) and 6 candidates (triangles) of the considered HARPS+CORALIE sample in the $m_2\sin{i}-\log{P}$ plane .}
  \label{HARPS_MM_m2sini-P_observations}
\end{figure}

\begin{figure}
  \resizebox{\hsize}{!}{\includegraphics{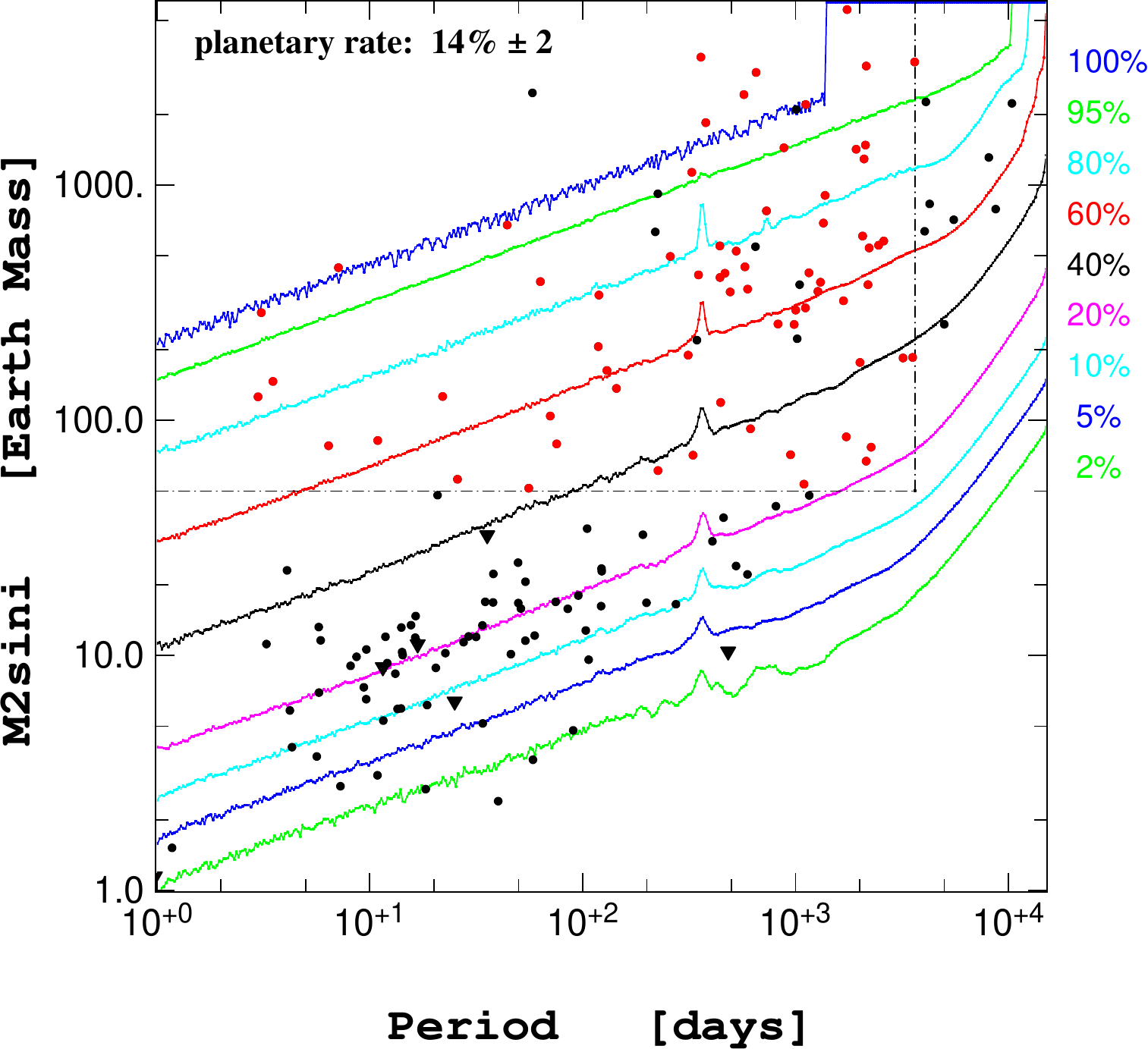}}
  \caption{Same as Fig.\,\ref{HARPS_MM_m2sini-P_observations} with detection probability curves superimposed. These detection probabilities are valid for the whole sample of 822 stars. After correcting for the detection bias, the fraction of stars with at least one planet more massive than 50\,M$_{\oplus}$ and with a period smaller than 10 years is estimated to be 14\,$\pm$\,2\,\%. The red points represent the planets which have been used to compute the corrected occurrence rate in the box indicated by the dashed line. The planets lying outside the box or being part of a system already taken into account are excluded; they are shown in black.}
  \label{HARPS_MM_m2sini-P_Msup50_Pinf10years}
\end{figure}

\begin{figure}
  \resizebox{\hsize}{!}{\includegraphics{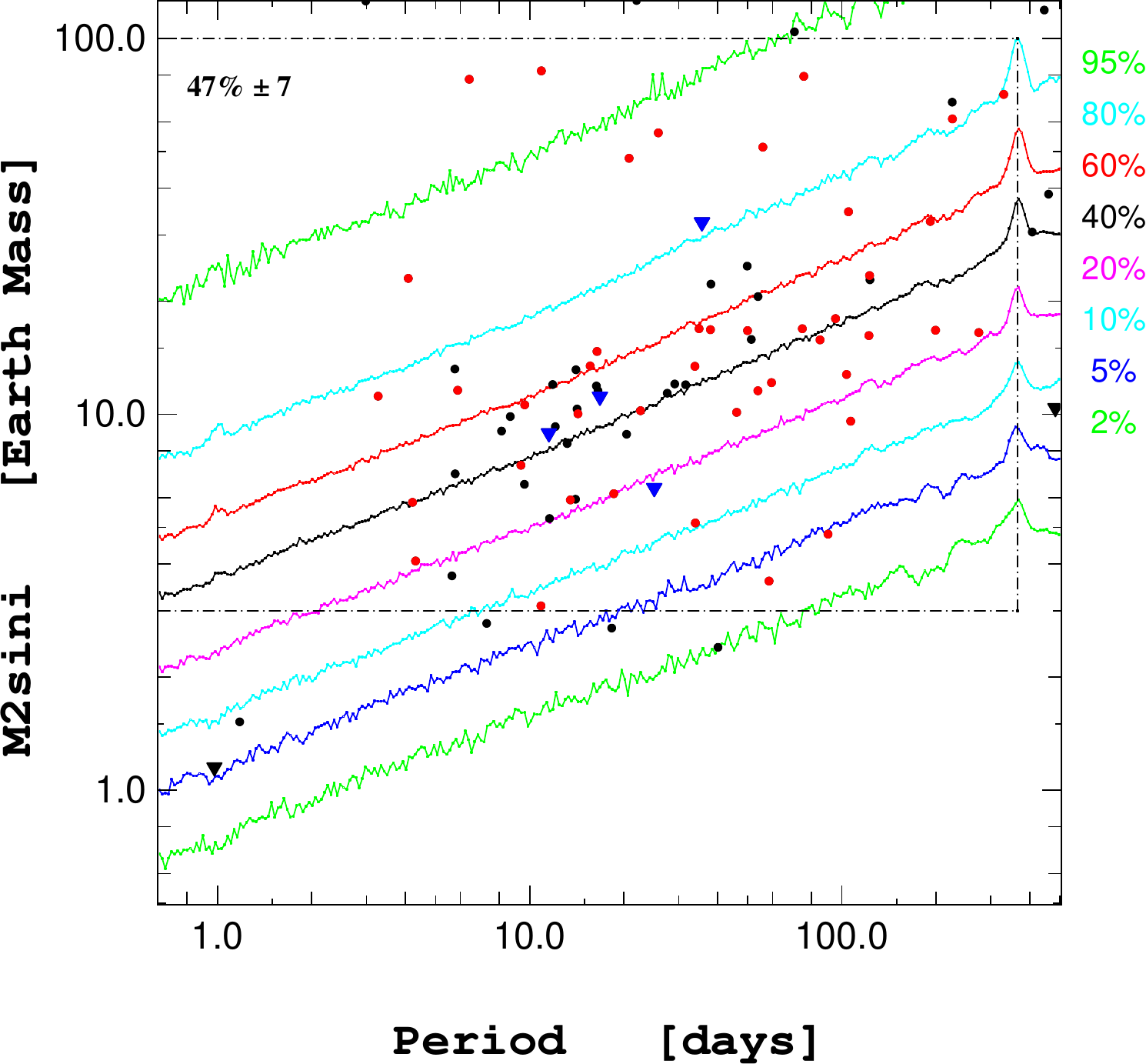}}
  \caption{Same as Fig.\,\ref{HARPS_MM_m2sini-P_observations} but only for the HARPS subsample. The occurrence rate of planetary systems in the limited region between 3 and 100\,M$_{\oplus}$, and with $P<1$~year, is $47\,\pm\,7$\,\%. Again, only the red dots and the blue triangles (candidates) have been considered for the computation of the occurrence rate.}
  \label{HARPS_MM_m2sini-P_3-100M_Pinf1year}
\end{figure}

\begin{figure}
  \resizebox{\hsize}{!}{\includegraphics{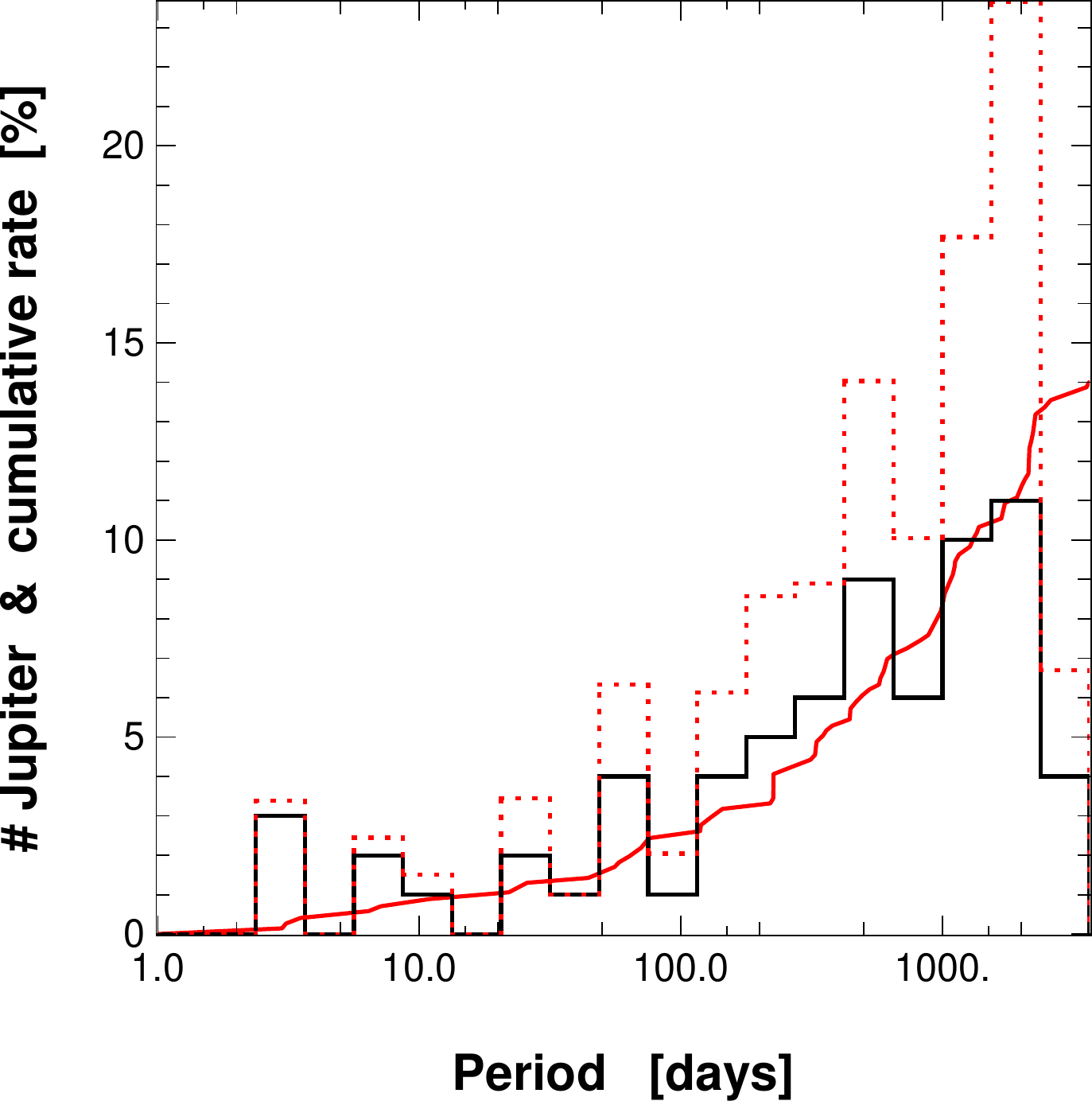}}
  \caption{Histogram of the planet frequency for planets with masses  ($m_2\sin{i}>50$\,M$_\oplus$). The occurrence rate for gaseous giant planets is strongly increasing with the logarithm of the period $\log{P}$.}
  \label{HARPS_MM_histo_P_Msup50M}
\end{figure}

%----------------------------------------------------------------
The $(m_2\sin{i},\log{P})$ diagram for the planetary systems and candidates detected in our survey is displayed in Fig.\,\ref{HARPS_MM_m2sini-P_observations}. Fig.\,\ref{HARPS_MM_m2sini-P_Msup50_Pinf10years} illustrates the same information, including lines of detection probability at various levels of completeness for the global sample. Several remarks can be made:
\begin{itemize}
\item[--] Objects with masses of about 1\,ms$^{-1}$ at very short period, and about 5\,ms$^{-1}$ at one year define a lower envelope of the displayed planets. The planets at this lower envelope are located in a region of very low-detection probability. The discovery of 9 planets close to the 2\,\% detection probability line was only possible thanks to an extremely large number of radial-velocity measurements of these stars. The detection of these planets is already a strong hint to the high occurrence rate of low-mass super-Earths.
\item[--] The density of massive super-Earths seems to exhibit a "mass-period" relation. The  extremum of the density appears to drift from about 6\,M$_\oplus$ at 10~days to more than 10\,M$_\oplus$ at 100 days. We do not believe that observation biases can explain the upper envelope of this feature relation, since heavy planets at sorter periods would be detected more easily (as indicated by the detection limits). Given the fact that the low-mass planets are all issued from the HARPS sample, it is interesting to consider the detection limits achieved for the 376 HARPS-stars only measured. The $(m_2\sin{i},\log{P})$ distribution for the HARPS subsample only is displayed on Fig.\,\ref{HARPS_MM_m2sini-P_3-100M_Pinf1year}. We note that the average detection probability of a Neptune-mass planet with a period of 100 days is still 60\,\%. The strongly decreasing density of super-Earths and Neptune-mass planets  (almost coincident with the red line corresponding to a detection of 60 percent) seems to be a highly significant feature.
\item[--] We do not have Neptune-mass planets (and objects up to 50\,M$_\oplus$) with periods larger than about 1000~days, despite the still non-neglectable detection probability in that domain of mass-period. For the moment this feature is not clearly understood. It could be the result of a detection bias related to the lack of measurements done for these low-amplitude and long-period objects, or simply due to the paucity of such planets beyond 1 or 2\,AU. For the moment we do not consider this feature as being significant.
\item[--] An "upper mass--$\log{P}$" relation is observed in the diagram, as already pointed out in \citep{Udry:2003}.  The diagram suggests that the upper mass for jovian planets on short periods (10~days) is close to one Jupiter and rises to 15~Jupiter masses at orbital periods of 10~Êyears.
\item[--] Not only the upper mass of planets is period dependent but the frequency of gaseous giant planets is strongly increasing with the logarithm of the orbital period, a feature already noticed in several giant planet searches \citep{Marcy:2005,Udry:2007a}. Figure\,\ref{HARPS_MM_histo_P_Msup50M} illustrates the histogram of the masses of planets with $m_2\sin{i}>50$\,M$_{\oplus}$ as a a function of orbital period. The observed distribution (continuous black line) as well as the bias-corrected distribution (dashed red line) illustrate the rising occurrence of gaseous planets with orbital periods. Setting the low-mass limit of the considered subsample to 100\,M$_\oplus$ does not change the general shape of the distribution.
\end{itemize}

%----------------------------------------------------------------
\subsection{Occurrence rate of planets as a function of planetary masses} 
%----------------------------------------------------------------
The estimation of the frequency of stars with a planet in a specific domain of the $m_2\sin{i}-\log{P}$ parameter space requires the correction of observations from the detection incompleteness. Actually as explained in Sect.\,\ref{sect_detlimit}, it is more taking into account the incompleteness determination in our occurrence estimate. In the case we consider a large zone in the $m_2\sin{i}-\log{P}$ plane, the presence of multi-planetary systems in the zone makes this operation ambiguous. For those systems, we have adopted the correction associated with the planet the most difficult to detect, as its higher statistical weight dominates the area.  
%We only provide the occurence rate of stars with planets above the lower detection limit, defined by the planet with the smallest detected radial-velocity amplitude (as a function of period). This lower limit  approximatively corresponds to the 2\,\% detection probability curve in Figs\,\ref{HARPS_MM_m2sini-P_observations} to \ref{HARPS_MM_m2sini-P_3-100M_Pinf1year}. 
We present in Table\,\ref{table_taux} the occurrence rate for different domains of masses and periods. \emph{We do not try to do any extrapolation of our results out of the range of the observed parameters, as e.g. towards smaller masses or longer periods.} For example, as we do not have any knowledge of the distribution of orbital periods of super-Earths above $\sim100$~days, we consider an extrapolation out to the habitable zone around the stars as being of no significance the time being. This is also true for the distribution towards very low masses, at any given period. An exponential extrapolation of the mass distribution towards very low mass planets is not justified if, for instance, a third, distinct population of Earth-mass planets is supposed, as suggested by Monte-Carlo simulations of planet population synthesis based on the core-accretion scenario \citep[e.g.][]{Mordasini:2009a,Mordasini:2009b}.

\begin{figure}
  \resizebox{\hsize}{!}{\includegraphics{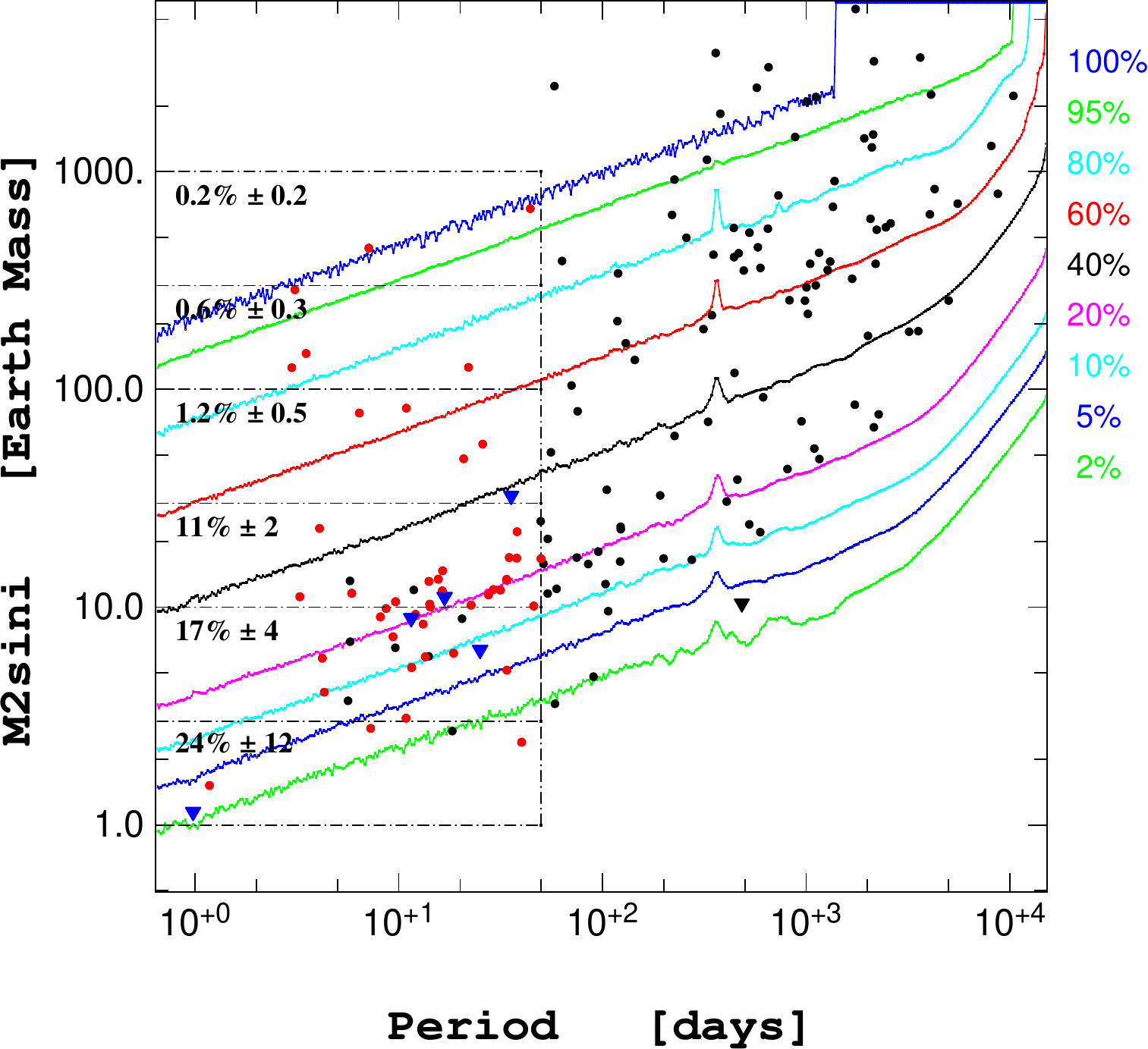}}
  \caption{Same as Fig.\,\ref{HARPS_MM_m2sini-P_Msup50_Pinf10years}. The dashed lines represent the boxes in which the occurrence rate is computed as defined by \citet{Howard:2010}. An additional box is shown for masses between 1 and 3\,M$_{\oplus}$.}
  \label{HARPS_MM_comparaison}
\end{figure}

Based on the $\eta$-Earth survey carried out at Keck observatory, \citet{Howard:2010} have derived an estimate of the occurrence rate of low-mass planets on tight orbits ($\leq50$\,days) as well as their mass distribution. The comparison of our results in the same range of parameter space with the $eta$-Earth estimate is given in Table\,\ref{table_taux_comparaison}. The results are also shown graphically in Fig.\,\ref{HARPS_MM_comparaison}, where the occurrence rate is indicated for the boxes defined by \citet{Howard:2010}. As a result of our much larger stellar sample, but also due to the better sensitivity of HARPS with regard to low-mass planets, the number of detected planets is significantly larger than the one issued from the Keck survey. This is specifically important for super-Earths and Neptune-mass planets (see Fig.\,\ref{HARPS_MM_K-P}. Our estimate for the occurrence rate of planetary systems is stunningly large. Table\,\ref{table_taux} lists the estimation of the frequency of planetary systems for different part of the $m_2\sin{i}-\log{P}$ plane. We should emphasize for example that 75\,\% of solar type stars have a planet with a period smaller than 10 years. This result is limited to the domain of detectable planets, and does not involve extrapolations out of that domain. Another interesting result is obtained for the low-mass planets on tight orbits: about half of solar-type stars is an host of this type of planets.

\begin{table*}
\caption{Occurrence frequency of stars with at least one planet in the defined region. The results for various regions of the $m_2\sin{i}-\log{P}$ plane are given.}
\label{table_taux}
\tabcolsep=3.4pt
\centering
\begin{tabular}{|c|c|c|c|c|}
\hline\hline
Mass limits  & Period limit  & Planetary rate based on   & Planetary rate             & Comments\\
                   &                    & published planets              & including candidates    &         \\
\hline
$>$\,50 M$_{\oplus}$  & $<$\,10 years & $13.9 \pm 1.7$\, \%  & $13.9 \pm 1.7$\,\%    & Gaseous giant planets\\ 
$>$\,100 M$_{\oplus}$ & $<$\,10 years & $9.7  \pm 1.3$\, \%  & $9.7  \pm 1.3$\,\%    & Gaseous giant planets\\ 
$>$\,50 M$_{\oplus}$  & $<$\,11 days  & $0.89 \pm 0.36$\,\%  & $0.89 \pm 0.36$\,\%   & Hot gaseous giant planets\\ 
Any masses            & $<$\,10 years & $65.2 \pm 6.6$\, \%  & $75.1 \pm 7.4$\,\%    & All "detectable" planets with $P\,<\,10$~years\\ 
Any masses            & $<$\,100 days & $50.6 \pm 7.4$\, \%  & $57.1 \pm 8.0$\,\%    & At least 1 planet with $P\,<\,100$~days\\ 
Any masses            & $<$\,100 days & $68.0 \pm 11.7$\,\%  & $68.9 \pm 11.6$\,\%   & F and G stars only\\
Any masses            & $<$\,100 days & $41.1 \pm 11.4$\,\%  & $52.7 \pm 13.2$\,\%   & K stars only\\
$<$\,30 M$_{\oplus}$  & $<$\,100 days & $47.9 \pm 8.5$\, \%  & $54.1 \pm 9.1$\,\%    & Super-Earths and Neptune-mass planets on tight orbits\\
$<$\,30 M$_{\oplus}$  & $<$\,50  days & $38.8 \pm 7.1 $\,\%  & $45.0 \pm 7.8$\,\%    & As defined in \citet{Lovis:2009}\\
\hline
\hline
\end{tabular}
\end{table*}

\medskip

\medskip

\begin{table*}
\caption{Comparison of detected planets detected and occurrence rate of the $\eta$\,Earth survey \citep{Howard:2010} and HARPS-CORALIE survey. The comparison is restricted to planets with orbital periods smaller than 50\,days.  $N_{1}$ stands for the numbers of detected planets, $N_{2}$ stands for the number of candidates, and $N_{3}$ represents the estimated occurrence rate of planets in the given mass range.}
\label{table_taux_comparaison}
\tabcolsep=3.4pt
\centering
\begin{tabular}{|c | c c c | c | c c c | c |}
\hline\hline
Mass range   & \multicolumn{4}{c|}{HARPS \& CORALIE survey} 
& \multicolumn{4}{c|}{$\eta$\,Earth survey} \\
\hline
      & \multicolumn{3}{c|}{Nb of planets}  & Planetary rate  & \multicolumn{3}{c|}{Nb of planets} & Planetary rate\\
M$_{\oplus}$ & $N_{1}$ & $N_{2}$ & $N_{3}$ &  $[\%]$  & $N_{1}$ & $N_{2}$ & $N_{3}$ & $[\%]$\\
\hline
3-10            & 19    & 2	 & 48.5    & $16.6 \pm 4.4$ & 5       & 3	& 10.2  & $11.8 \pm 4.3$ \\
10-30          & 25    & 1	 & 20.6    & $11.1 \pm 2.4$ & 4       & 1	& 4.6    & $6.5  \pm 3.0$ \\
30-100        & 5    & 1    & 4.6  & $1.17 \pm 0.52$   &   2       & 	&	     & $1.6  \pm 1.2$  \\
100-300     & 4     & 0    & 0.8  & $0.58 \pm 0.29$   &    2       & 	&	     & $1.6  \pm 1.2$  \\
300-1000   & 2     & 0    & 0      & $0.24 \pm 0.17$  &    2       & 	&	     & $1.6  \pm 1.2$  \\
\hline
\hline
\end{tabular}
\end{table*}

\medskip

Despite the significant size of our sample of 822 stars, the number of hot Jupiters is quite small (5 planets with $P<11$\,days). The estimated frequency,  0.9 $\pm$ 0.4\% is compatible with previous estimates. (For a more complete discussion of the CORALIE-alone sample we refer to Marmier et al. (in prep). We know that the mass distribution of planets hosted by G dwarfs is different from its equivalent for planets hosted by M dwarfs \citep{Bonfils:2011}. Despite the rather limited range of stellar masses in our sample, we have tried a comparison of the $(m_2\sin{i}-\log{P})$ distribution for dwarf stars of spectral type F and G versus the distribution for K dwarfs. The observed difference of planetary rate (for periods smaller than 100~days) for the two spectral types is however not significant (Table\,\ref{table_taux}).

%----------------------------------------------------------------
\subsection{The mass distribution} 
\begin{figure}
  \resizebox{\hsize}{!}{\includegraphics{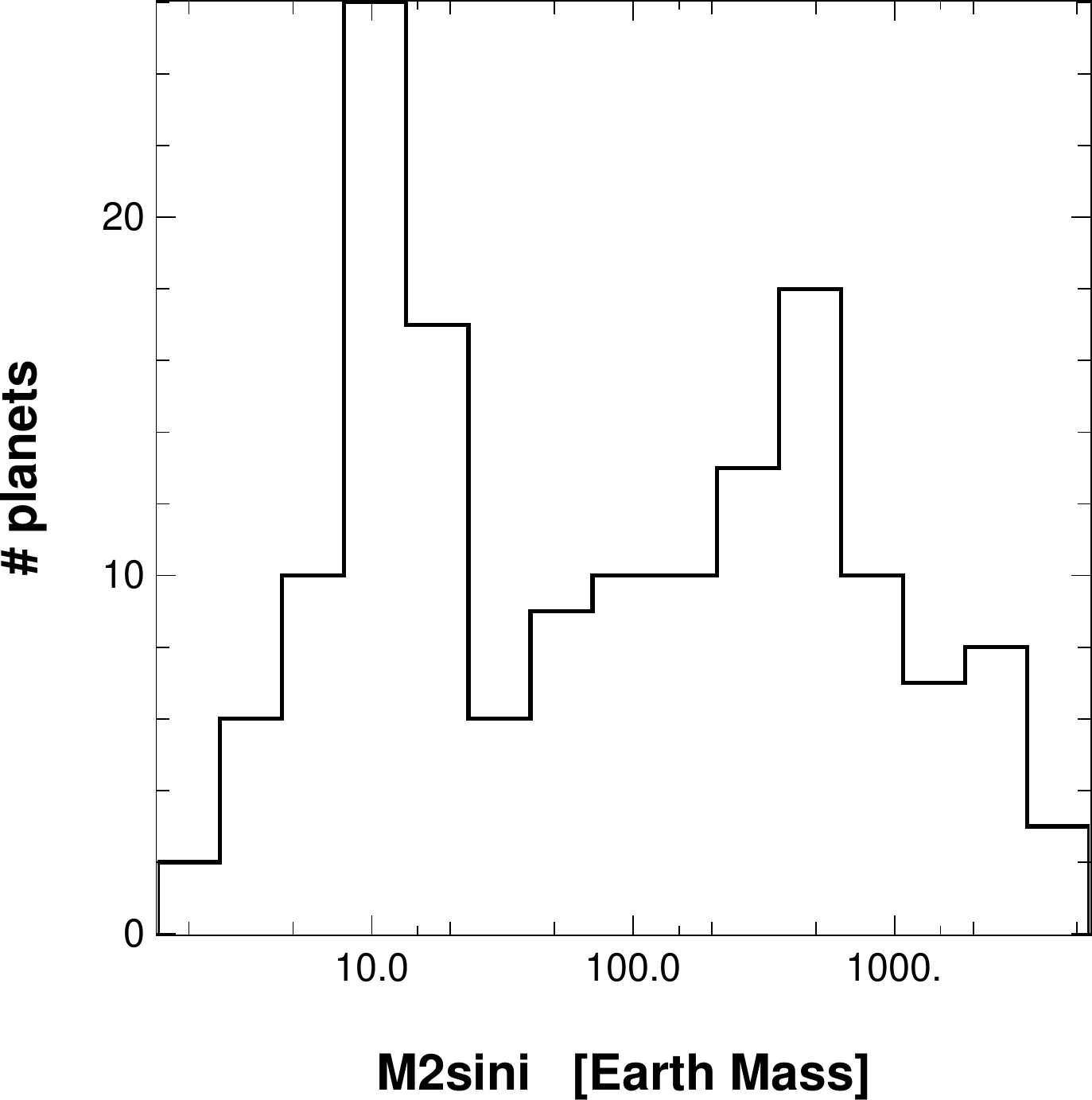}}
  \caption{Observed mass histogram for the planets in the combined sample. Before any bias correction, we can already notice the importance of the sub-population of low-mass planets. We also remark a gap in the histogram between planets with masses above and below $\sim$30\,M$_\oplus$.}
  \label{HARPS_MM_histo_m2sini_completesample}
\end{figure}

\begin{figure}
  \resizebox{\hsize}{!}{\includegraphics{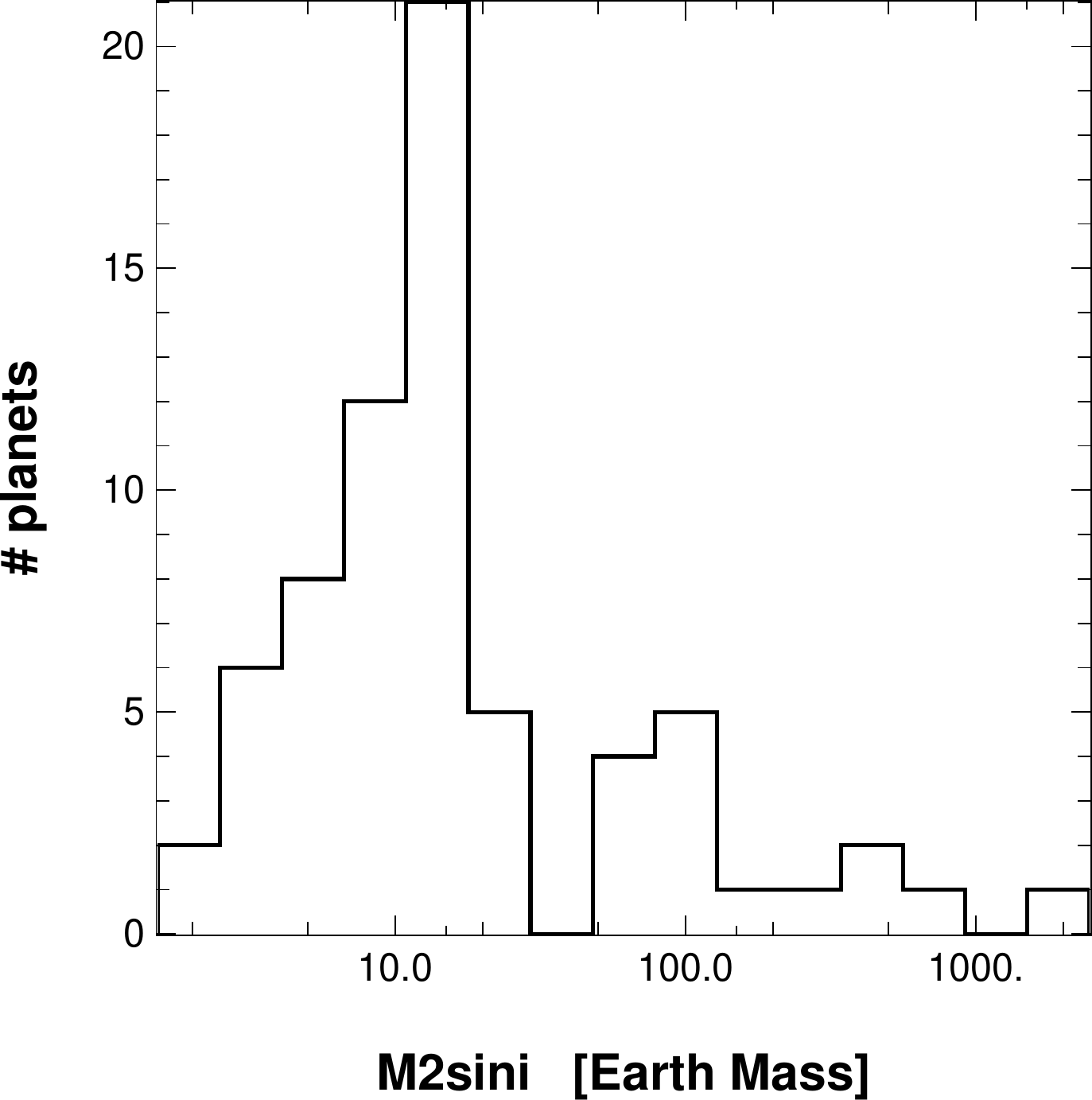}}
  \caption{Same as Fig.\,\ref{HARPS_MM_histo_m2sini_completesample} but for planets with periods smaller than 100~days. We see the dominance of low-mass planet with short orbital periods.}
  \label{HARPS_MM_histo_m2sini_Pinf100d}
\end{figure}

\begin{figure}
  \resizebox{\hsize}{!}{\includegraphics{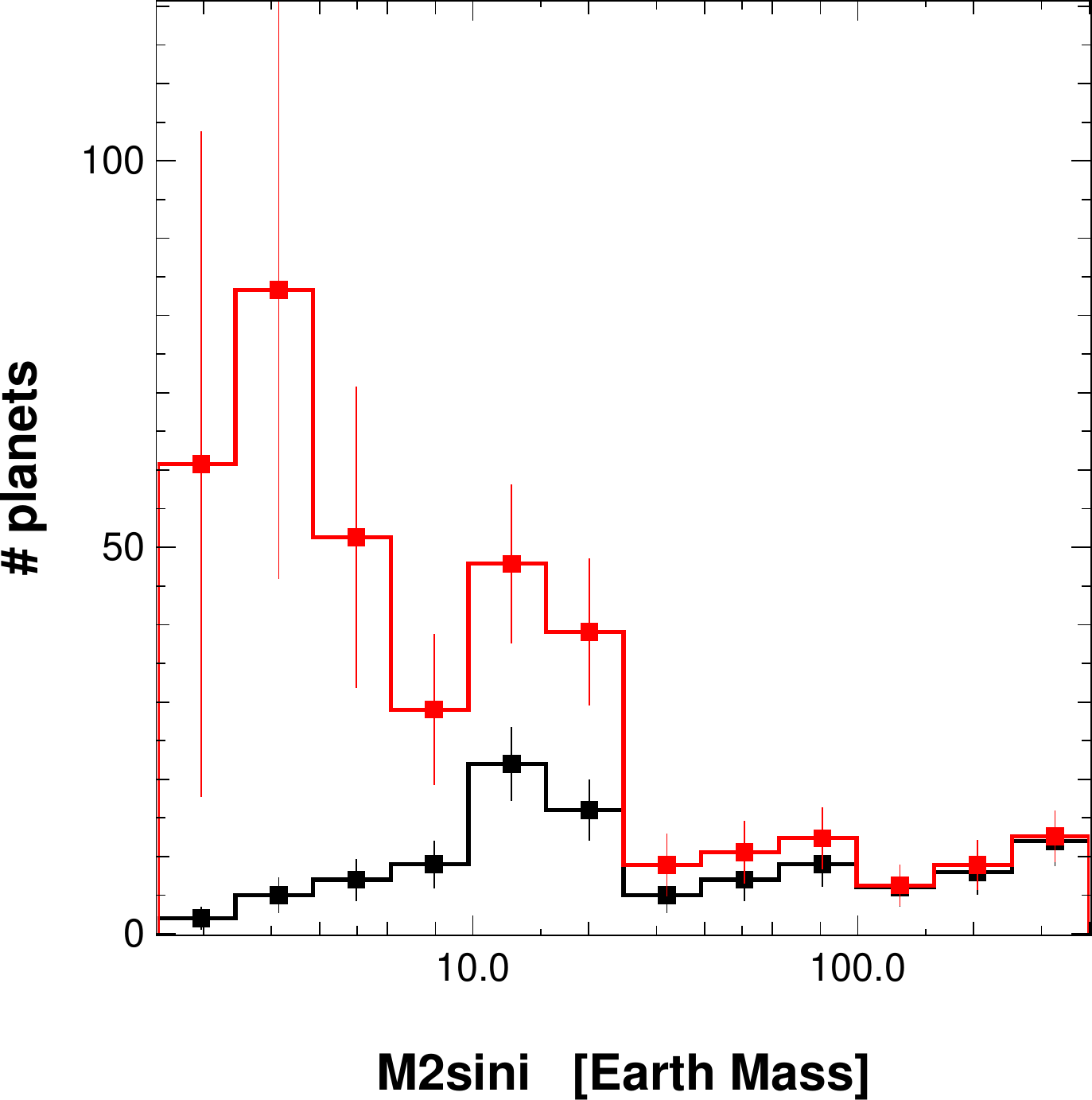}}
  \caption{Histograms of planetary masses, comparing the observed histogram (black line) and the equivalent histogram after correction for the detection bias (red line).}
  \label{HARPS_MM_histo_mass_biascorrected}
\end{figure}

%----------------------------------------------------------------
On Fig.\ref{HARPS_MM_histo_m2sini_completesample} we have plotted the histogram of masses of the planets detected in our sample. We observe a drastic decline of the observed mass distribution from about 15 to 30\,M$_\oplus$. If we limit the range of orbital periods and only consider planets with $P<100$\,days (Fig.\,\ref{HARPS_MM_histo_m2sini_Pinf100d}), a region where the detection bias are not too important for low-mass planets, we immediately observe the preponderant importance of the sub-population of super-Earths and Neptune-mass planets in that domain of periods.  After  correction of detection biases (Fig.\,\ref{HARPS_MM_histo_mass_biascorrected}), we see even more clearly the importance of the population of low-mass planets on tight orbits, with a sharp decrease of the distribution between a few Earth masses and $\sim 40$\,M$_{\oplus}$. We note that the planet population synthesis models by \citet{Mordasini:2009a} predicted such a minimum in the mass-distribution at precisely this mass range. They also pointed out that a radial-velocity measurement precision of about 1\,ms$^{-1}$ was required in order to detect this minimum. In the framework of the core accretion model, this can be understood by the fact that this mass range corresponds to the runaway gas accretion phase during which planets acquire mass on very short timescales. Therefore, unless timing is such that the gaseous disk vanishes at this moment, forming planet transits quickly through this mass range and the probability to detect these types of planets is reduced correspondingly. In Fig.\,\ref{HARPS_MM_histo_mass_biascorrected} the importance of the correction of the detection biases below 20\,M$_\oplus$ is only the reflection of the present observing situation for which only a limited fraction of the sample has benefited from the large enough number of HARPS measurements, required to detect small-mass objects. Part of this correction is also related to the growing importance of the $\sin{i}$ effect with decreasing masses.

%----------------------------------------------------------------
\subsection{The period distribution of Super-Earth and Neptune-mass planets} 
\begin{figure}
  \resizebox{\hsize}{!}{\includegraphics{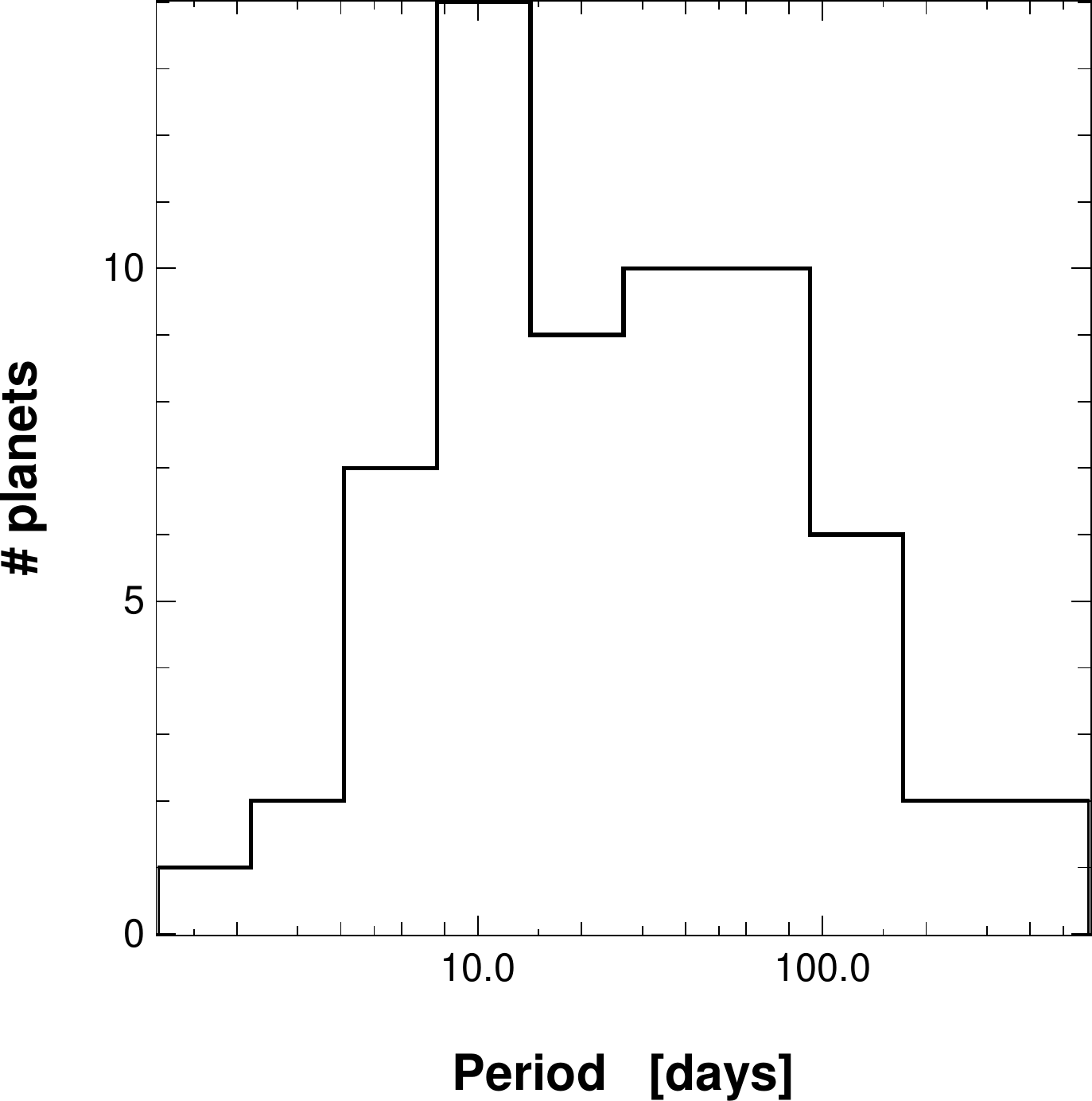}}
  \caption{Observed period distribution for low-mass planets ($m_2\sin{i}<30$\,M$_\oplus$)}
  \label{HARPS_MM_histo_P_massinf30M}
\end{figure}

\begin{figure}
  \resizebox{\hsize}{!}{\includegraphics{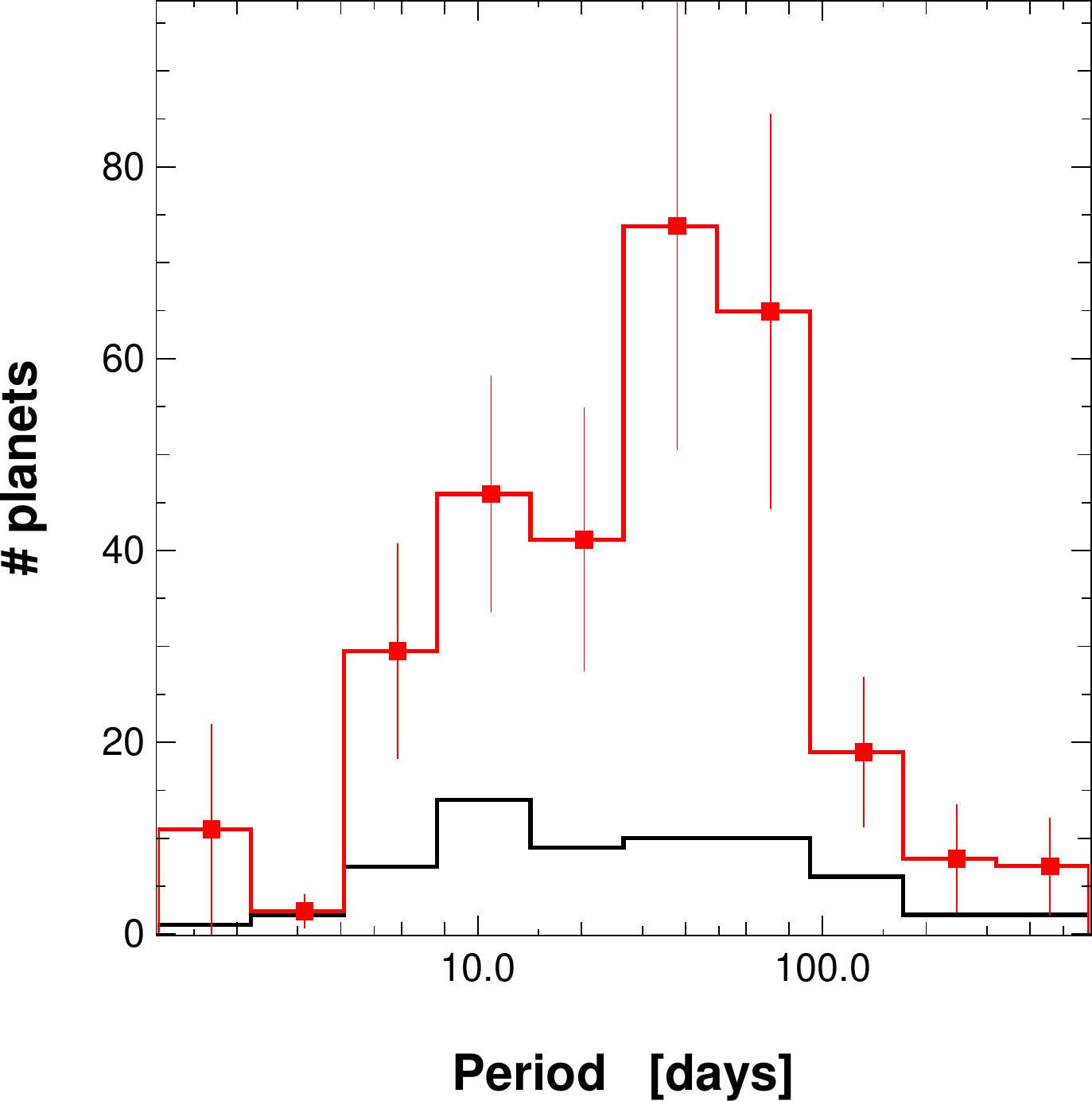}}
  \caption{Observed period distributions for low-mass planets ($m_2\sin{i}<30$\,M$_\oplus$) before (black histogram) and after (red histogram) correction for the detection bias. Most of the low-mass planets are confined on short period orbits. The mode of the distribution appears just over 40~days.}
  \label{HARPS_MM_histo_P_massinf30M_biascorr}
\end{figure}

%----------------------------------------------------------------
The observed distribution of orbital periods for planets less massive than 30\,M$_\oplus$ is illustrated in Fig.\ref{HARPS_MM_histo_P_massinf30M}. In Fig.\ref{HARPS_MM_histo_P_massinf30M_biascorr}, the same distribution is reproduced with a black histogram, to be compared with the histogram after correction for detection incompleteness (red histogram). In agreement with Kepler's preliminary findings \citep{Borucki:2011}, the sub-population of low-mass planet appears mostly confined to tight orbits. The majority of these low-mass planets have periods shorter than 100\,days. Low-mass planets on longer periods are of course more affected by detection limits, this is however, at least partly, taken into account in our bias estimate and correction. We conclude that this feature must be real.
\medskip

%----------------------------------------------------------------
\subsection{Orbital eccentricities of Super-Earth and Neptune-type planets} 
\begin{figure}
  \resizebox{\hsize}{!}{\includegraphics{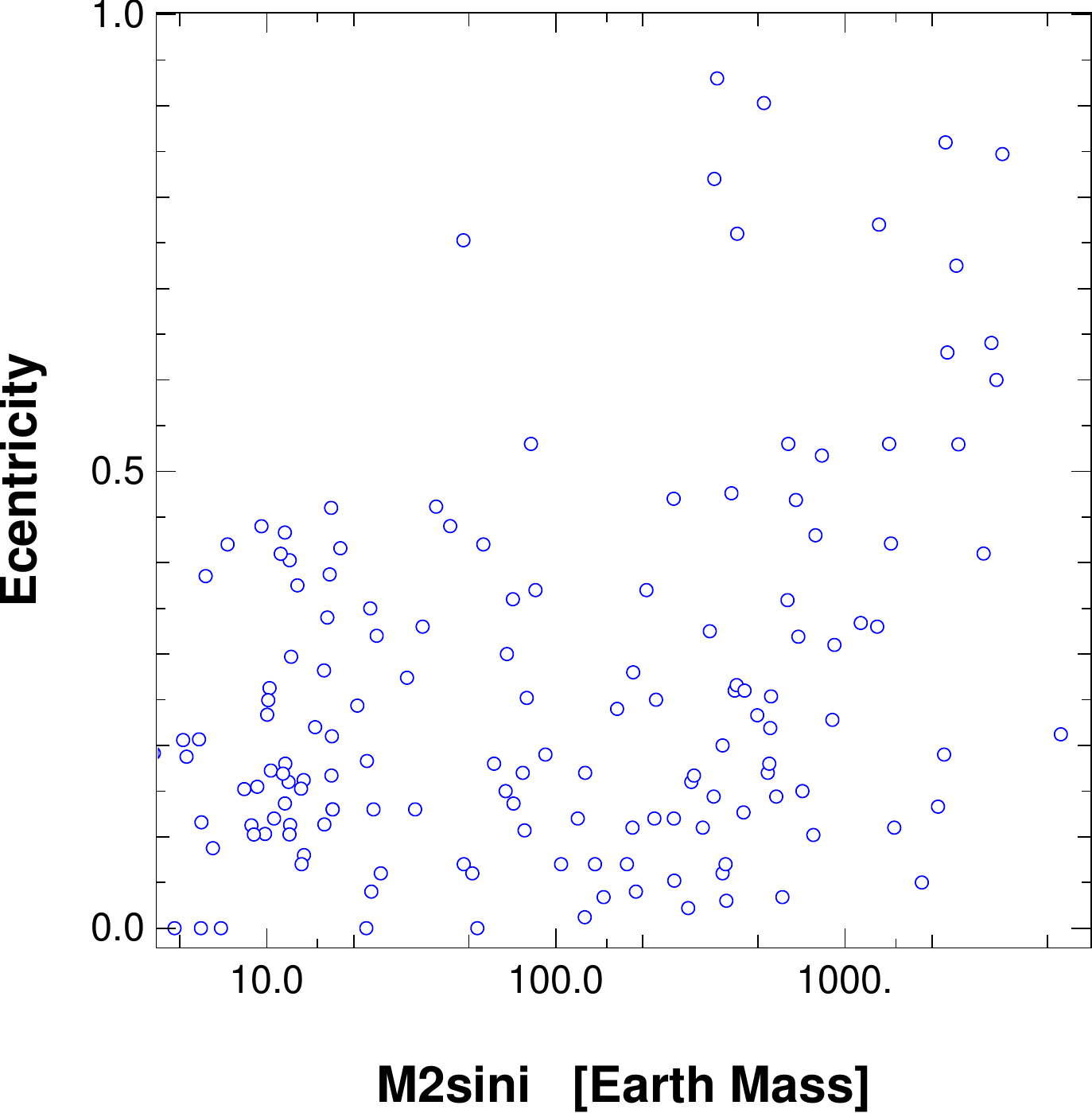}}
  \caption{Mass-eccentricity diagram for the planets in the combined sample.}
  \label{HARPS_MM_eccentricity}
\end{figure}

%----------------------------------------------------------------
Figure\,\ref{HARPS_MM_eccentricity} displays the orbital eccentricities as a function of the planetary mass. We can remark the very large scatter of orbital eccentricities measured for gaseous giant planets, some of them having eccentricities as large as 0.93. Such very large eccentricities are not observed for planets with masses smaller than about 30\,M$_\oplus$ for which the most extrem values are limited around 0.45. 
For low-mass planets the estimation of small orbital eccentricites of the best keplerian fit is biased. For the moment, the eccentricities below 0.2 (and small masses) have to be considered with caution .  

%-----------------------------------------------------------------
\subsection{Fraction of multiplanetary systems with low mass planets}
%------------------------------------------------------------------

For systems with planets less massive than 30\,M$_\oplus$, the fraction of multi-planetary systems is extremely high. For the 24 concerned systems this fraction exceeds 70\,\%. It is tempting to have a rate of multi-planetary systems hosting at least one gaseous giant planets. Unfortunately, the optimum observing strategy needed to detect low-mass planets has not been applied to every stars with giant planets in past. Presently, we observe a multi-planetary rate of only 26\,\% for these planets. This point will have to be revisited with additional precise velocities. For the formation of planetary systems, the existence of systems with one gaseous giant planet with large period and a low-mass planet on a tight orbit could be of interest. A few systems having these characteristics can be listed: HD\,10180, HD\,11964, HD\,134060, HD\,160691, HD\,181433, HD\,204313, and HD\,215456.

%----------------------------------------------------------------
\section{Host star metallicities as a function of planetary masses} 
%----------------------------------------------------------------
The occurrence rate of giant gaseous planets strongly correlates with the host star metallicity. Large unbiased studies have provided a robust and well-defined relationship between the frequency of gaseous giant planets and the metallicity of their host star \citep[e.g.][]{Santos:2001, Santos:2004a, Fischer:2005}. The lack of correlation between the overabundance of heavy elements and the mass of the convective zone of the star \citep{Santos:2004a} is seen as a strong argument in favour of an origin of the planet-metallicity correlation linked with the primordial abundance of the molecular cloud. 

\begin{figure}
  \resizebox{\hsize}{!}{\includegraphics{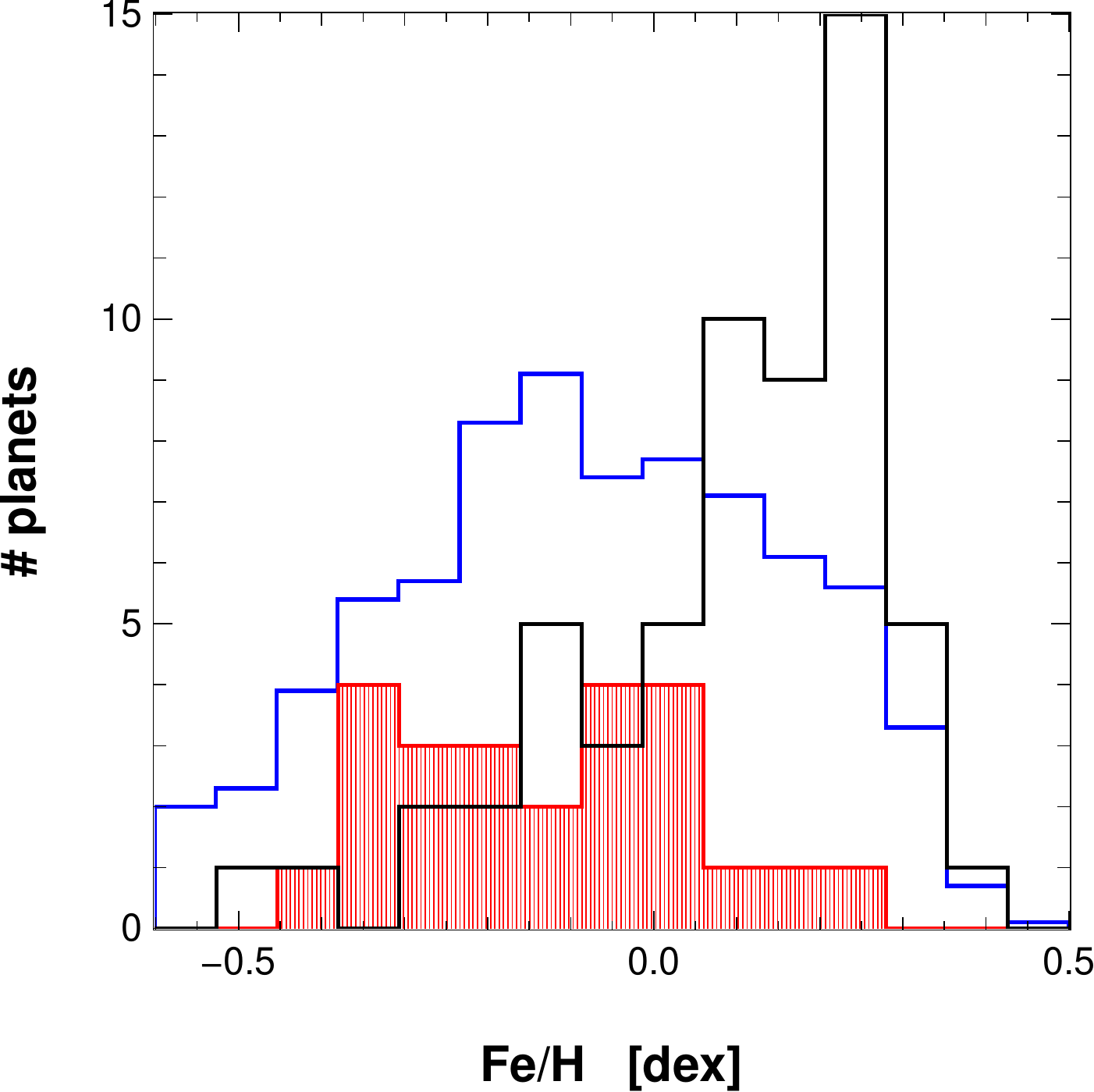}}
  \caption{Histograms of host star metallicities ([Fe/H]) for giant gaseous planets (black), for  planets less massive than 30\,M$\oplus$ (red), and for the global combined sample stars (blue). The latter histogram has been multiplied by 0.1 for visual comparison reason.}
  \label{HARPS_MM_FeH_Minf30M_all}
\end{figure}

\begin{figure}
  \resizebox{\hsize}{!}{\includegraphics{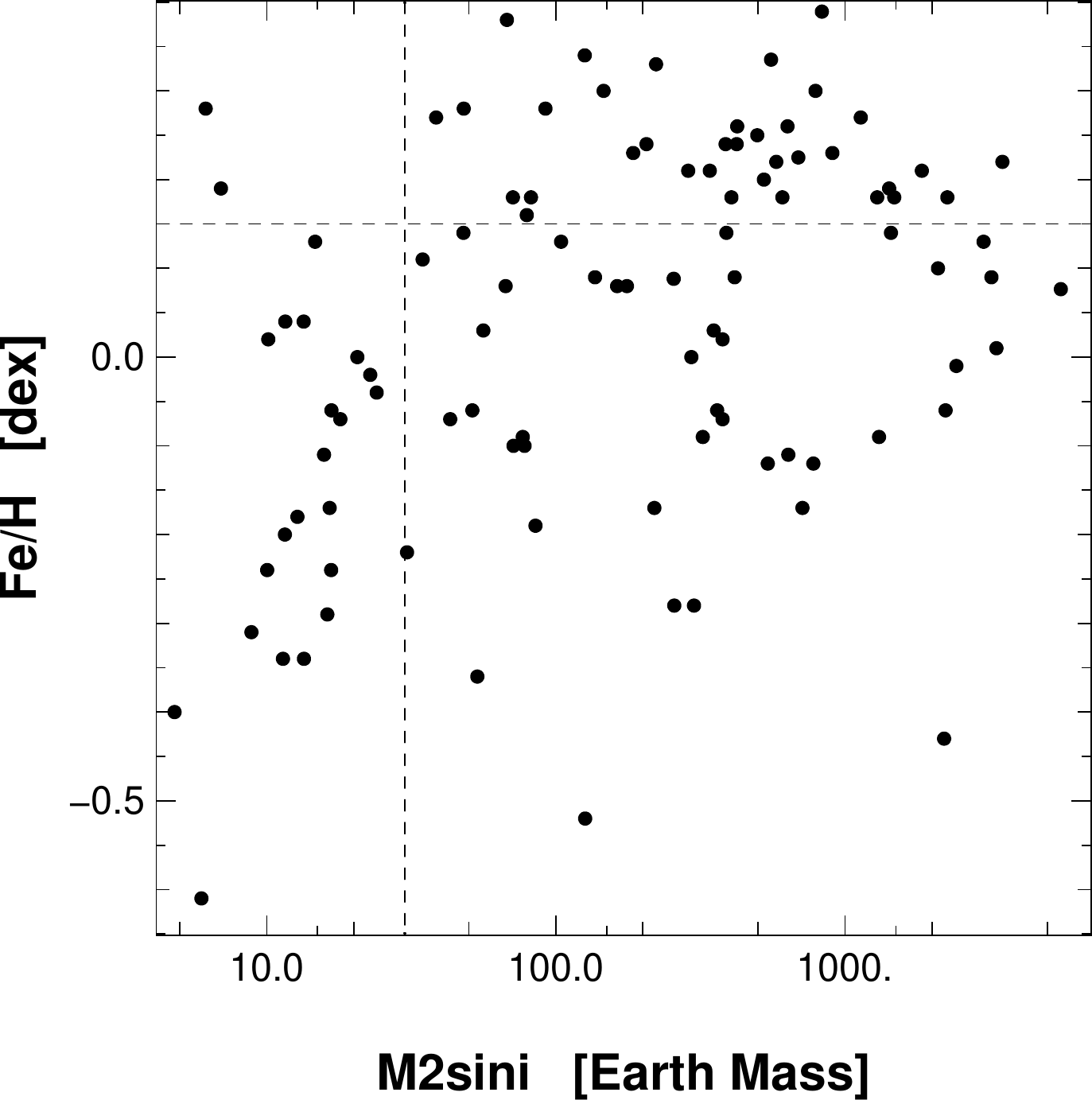}}
  \caption{Estimation of the planetary-mass limit between the two regimes for the metallicity dependance of host stars. A vertical line at 30\,M$_\oplus$ distinguishes the two populations. We should note that such a limit also corresponds to the gap in the mass distribution (see Fig.\ref{HARPS_MM_histo_m2sini_completesample} and \ref{HARPS_MM_histo_mass_biascorrected}). On the right side of the vertical line we do not observe significant changes of the metallicity distribution above 30\,M$\oplus$. We remark that stars with metallicity exceeding 0.15 are for their huge majority associated with planets more massive than 30\,M$_\oplus$. }
  \label{HARPS_MM_FeH-mass_30M}
\end{figure}

For lower-mass planets, already after the very few first detections \citep{Udry:2006} suggested the absence of correlation between the host star metallicity and the presence of low-mass planets. This first claim was later confirmed by \citep{Sousa:2008} on a larger but still limited sample. The present analysis of a statistically well-defined, much larger sample of low-mass planets, offers the opportunity to have a much more robust insight in the relation between planet occurrence frequency, host star metallicity, and planet mass. For the analysis, multi-planetary systems will be only characterized by their most massive planet. In Fig.\,\ref{HARPS_MM_FeH_Minf30M_all}, we can compare the histograms of host-star metallicities for planets with masses smaller than 30\,M$_\oplus$ and its equivalent for gaseous giant planets. On the same plot, the metallicity distribution for the stars in the global combined sample is illustrated for an enlightening comparison. It is however not clear whether we observe a discontinuity in the-host star metallicity distributions with an increasing planetary mass. On Fig.\ref{HARPS_MM_FeH-mass_30M}, the ([Fe/H], $m_2\sin{i}$) diagram can be used to set a limit between the two regimes of host star metallicity (if such a limit proves to be meaningful!). On the diagram, such a limit can be set at about 30-40\,M$_\oplus$. At the exception of a single star, all the stars hosting planets less massive than 40\,M$_\oplus$ have metallicities below [Fe/H]\,=\,0.20. This is well in contrast with the situation for stars hosting more massive planets. Interestingly this mass corresponds about to local the minimum in the mass distribution between Neptunes and gaseous giants (see e.g. Fig.\,\ref{HARPS_MM_histo_m2sini_completesample} or Fig.\,\ref{HARPS_MM_histo_m2sini_Pinf100d}). 

The correlation between the occurrence of giant planets and the metallicity of the host star (i.e. the metallicity of the material in the proto-planetary disc) is a natural outcome of the core accretion planet formation theory. In this paradigm, massive gaseous planets form by runaway gas accretion onto cores exceeding a critical mass, typically of the order of $10-20$\,M$_\oplus$. The gas accretion from the disc goes on until the disc vanishes, typically after a few million years. Hence, the sooner in the evolution of the disk a critical-mass core can form, the larger the amount of gas that will still be available for accretion.  A high metallicity (interpreted as a large dust-to-gas ratio in the models) and/or massive discs favors the early growth of such critical cores. Conversely, lower-mass planets that do not accrete significant amount of gas, can grow their cores over a longer timescale and therefore do not depend as critically upon the metallicity. These effects have been born out in the population synthesis models by \citet{Mordasini:2011}.

%----------------------------------------------------------------
\section{Planets in the habitable zone of solar-type stars} 
%----------------------------------------------------------------

All the very specific properties of the population of low-mass planets (super-Earths and Neptune-mass planets) are of special interest for constraining the formation of planetary systems. In addition to this,  the surveys targeting low-mass planets, in the same way as the parallel efforts aiming to increase the precision of spectrographs optimized for Doppler measurements are pursuing a still more difficult challenge, the detection of Earth twins, rocky planets orbiting stars in the so-called habitable zone, and possibly around stars as close as possible to the Sun. This last condition is of special importance for future experiments aiming at the spectroscopic follow-up of the planet, in order to e.g. characterize its atmosphere. For this long-term goal we would like to contribute to an "input catalogue" with a significant number of entries. Would Doppler spectroscopy have a chance to fulfill such an ambitious goal?

At present, already 2 super-Earths located in the habitable zone of their host star have been detected with the HARPS instrument. The first one, Gl\,581\,d is part of a multi-planetary system hosting 4 low-mass planets. Gl\,581\,d has been detected by \citet{Udry:2007b}. As a result of the aliasing with the sideral year, its orbital period had to be corrected later \citep{Mayor:2009b}. The minimum mass and orbital period of Gl\,581\,d are 7\,M$_\oplus$ and 66~days. Despite its rather short period the planet appears as the first super-Earth discovered in the habitable zone due to the very late spectral type and low mass of Gl\,581 (M5V, 0.3\,M$_\oplus$). Models of the atmosphere of Gl\,581\,d have demonstrated the possibility of habitability \citep{Kaltenegger:2011a,Wordsworth:2011,Hu:2011}. The discovery of another super-Earth in the habitable zone of the same star (G\,581\,g) was claimed by \citep{Vogt:2010}. Statistical reanalysis of the published velocity data could unfortunately not confirm the detection \citep{Andrae:2010,Gregory:2011,Tuomi:2011}. Doubling the number of available precise HARPS measurements \citet{Forveille:2011} ruled out the existence of Gl\,581\,g. 

A new program to explore the possibility to detect habitable planets orbiting solar-type stars with HARPS started 2 years ago \citep{Pepe:2011}. A measurement strategy is applied to limit the influence of stellar noises (acoustic and granulation noises) and observations are carried out as often as possible, ideally over several years, in order to extract a potential planetary signal possibly hampered by lower frequency noise sources (spots etc). This observing strategy is quite demanding in terms of telescope time, and forced us to limit the size of this precursor sample to only 10 solar-type stars. These 10 stars where selected according to their proximity ($<16$\,pcs) and for having a low level of chromospheric activity. After only 2 years, already 5 low-mass planets have been discovered orbiting three stars among the ten \citep{Pepe:2011}. We should first mention a  3-planet system hosted by the metal deficient ([Fe/H]= -0.40) star G4V HD\,20794. For these three planets the detected very-low radial-velocity amplitudes (0.83, 0.56 and 0.85 ms$^{-1}$) are already impressively small and correspond to masses of (2.7,  2.4 and 4.8\,$M_{\oplus}$), respectively. Another low-mass planet has also been detected \citep{Pepe:2011}; it is hosted by the metal-deficient ([Fe/H] = - 0.33) K5V star HD\,85512.  Once again, the velocity amplitude is smaller than the meter per second (0.77\,$\pm$\,0.09\,ms$^{-1}$). With an orbital period of 58.4 days and a minimum mass of 3.6\,M$_{\oplus}$, HD\,85512\,b seems to be located inside, but close to the inner boundary, of the habitable zone of this K5 star \citep{Kaltenegger:2011b}.

\begin{figure}
  \resizebox{\hsize}{!}{\includegraphics{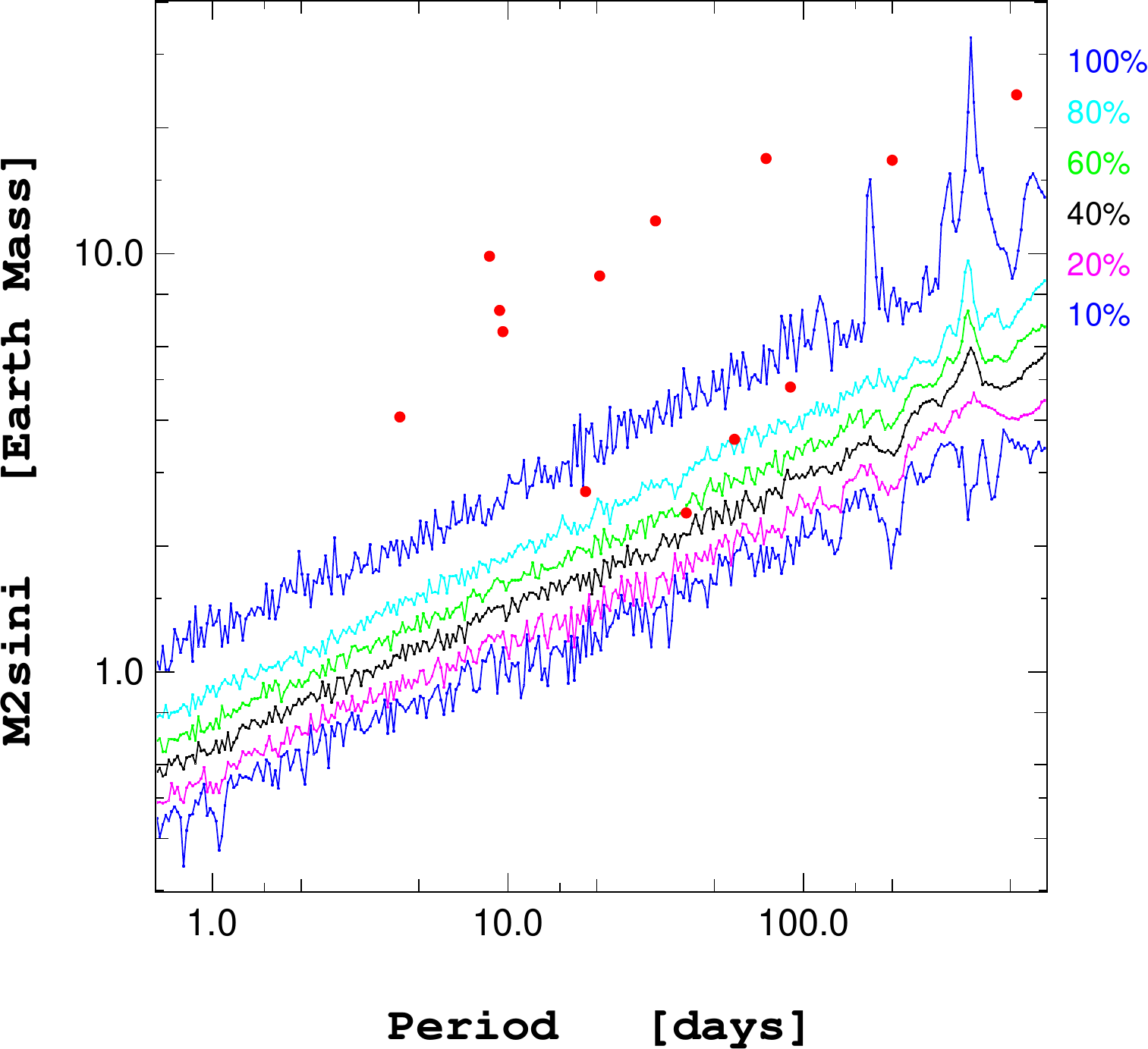}}
  \caption{The figure illustrates the limits of detection for very low-mass planets based on only the 10 stars of the HARPS sample with more than 165 HARPS measurements per star. Mote that 29 planets have been discovered orbiting these ten stars!  }
  \label{HARPS_MM_10stars}
\end{figure}

Let us notice the proximity to the solar system of Gl\,581 (6.06 pcs)  and HD\,85512 (11.15 pcs), and their age very similar to the Sun's one (5.8 and 5.6 Gyr). The detection of HD\,85512\,d is close to the HARPS limit of detectability but demonstrates the possibility of detecting super-Earths in the habitable zone of solar-type stars. To evaluate the sensitivity of Doppler spectroscopy to detect super-Earths in the habitable zone we have selected the 10 stars measured with HARPS and having the largest number of measurements. When multiple measurements are done on the same night, with the purpose to reduce the granulation noise influence, these measurements are counted for a single data point. All these stars have been measured more than 165 times during several years. Altogether 29 planets have been discovered orbiting these ten stars. We have evaluated the detection limits for these 10 stars, and sketched them in the $m_2\sin{i}-\log{P}$ plane (see Fig.\ref{HARPS_MM_10stars}). Consodering the actual calendar of measurements of the ongoing HARPS survey, if the number of measurements is large enough (let say larger than 165 per star), we obtain a detection sensitivity close to 100\,\% for a 10\,M$_\oplus$ super-Earth up to orbital periods of one year. The corresponding detection probability is still close to 20\,\% for a 3\,M$_\oplus$ planet. 

A new spectrograph called ESPRESSO is presently developed for ESO's VLT at the Paranal Observatory. We can also mention the development of a northern copy of HARPS to be installed on the TNG (3.5-m telescope) at La Palma, Canary Islands. This project should contribute to the radial velocity measurements and mass determination of Kepler mission's planet candidates. Part of the HARPS-N observing time will also be devoted to the detection of Earth twins orbiting solar-type stars in the solar vicinity. We can be confident that in the coming ten to twenty years we should have, by the radial-velocity measurements, an important contribution to a first list of solar-type stars with habitable planets and an estimation of their frequency.

%----------------------------------------------------------------
\section{Conclusions} 
%----------------------------------------------------------------
The (still continuing) HARPS survey has already allowed a detailed study of statistical properties of planetary systems over the mass domain from a few Earth masses to several Jupiter masses. This survey provides information
over three orders of magnitude of planetary masses. Thanks to the sub-meter-per-second precision of HARPS and the significant number of observing nights dedicated to this survey over the last 8 years, exceptional results have been obtained. The overwhelming importance of the population of low mass planets on tight orbit has been quantified. 
About 50\% of solar-type stars host planets with mass lower that 30\,M$_{\oplus}$ on orbits with period shorter than 100\,days. The lower limit for this frequency value is set by a residual detection bias below two to five Earth-masses at periods from few to 100\,days. Quite obviously any extrapolation to lower masses in that range of periods would further increase this occurrence rate. On the other hand, because of the possible existence of a population of solid planets with rising density below a few Earth-masses \citep{Mordasini:2009a} any extrapolation to lower masses is highly speculative.

The time being, the presented mass distribution indicates that we are in presence of two populations of planets. They are separated by a "gap" in the mass distribution at about 30 Earth-masses, which cannot be the consequence of a detection bias. These populations have quite different characteristics:

\begin{itemize}
\item The population of gaseous giant planets (GGP) with masses above 50 Earth-masses covers the range of periods from days to several thousands of days, but the frequency is rising with the logarithm of the period $\log P$ (Fig.\,\ref{HARPS_MM_m2sini-P_3-100M_Pinf1year}). These planets exhibit an upper-mass limit which increases with the orbital period (Fig.\,\ref{HARPS_MM_m2sini-P_Msup50_Pinf10years}). The occurrence rate of GGP is about 14\% ($m_2 \sin i > 50$\,M$_{\oplus}$, $P < 10$\,years). Furthermore, the occurrence rate of GGP is strongly correlated with the host star metallicity (see black histogram in Fig.\,\ref{HARPS_MM_histo_mass_biascorrected}). In a systematic survey to search for planets orbiting metal deficient stars, \citet{Santos:2011} have detected only three GGP among 120 observed stars. All the three GGP have host star metallicities close to -0.5\,dex, which was the upper limit set to the metallicity of the host stars in that survey. It seems that the observations set some lower limit to the host star metallicity (and supposingly to the accretion disk) to allow for the formation of GGP. Finally, we observe in this sample that the distribution of orbital eccentricities exhibits a large scatter and may attain values even larger than 0.9.

\item The population of Super-Earth and Neptune-type planets (SEN) with masses less than about 30 to 40 Earth-masses behaves quite differently. About 50 percents of solar-type stars host at least one super-Earth or Neptune. The mass distribution is strongly decreasing from 15 to 30\,M$_{\oplus}$ despite the fact that they would be detected much more easily (Fig.\,\ref{HARPS_MM_histo_P_Msup50M} and Fig.\,\ref{HARPS_MM_histo_m2sini_completesample}). Also, the SEN population seems to prefer intermediate orbital periods from 40 to 80\,days once corrected for (Fig.\,\ref{HARPS_MM_histo_mass_biascorrected}), their frequency decreasing on both ends toward shorter and longer periods. Opposite to GGP, the occurrence rate of SEN does not exhibit a preference for metal rich host stars. The difference between both populations of planets is striking (see the host-star metallicity histogram in Fig.\,\ref{HARPS_MM_FeH_Minf30M_all}). The median metallicity for solar-type stars host of SEN planets is close to Fe/H = -0.1. A significant number of SEN have host stars as metal deficient as -0.4\,dex. It must be recalled here that only the most massive planet of a multi-planetary system is used for that comparative study of GGP versus SEN systems. A systematic search is continuing with HARPS (Pi N.Santos) to explore the occurrence of SEN hosted by very metal-deficient stars. At present the lower limit of host-star metallicity needed to form a planetary system remains unknown. A short final word should be spent on eccentricities of SEN. They are clearly different from zero in many cases, but in contrast to GGP they are limited to more modest values up to 0.45.
\end{itemize}

It is interesting to note that our estimation for the occurrence rate of low-mass planets with periods shorter than 100\,days is larger than 50 percents. This value is remarkably different from the estimation resulting from the Kepler mission. One simple explanation could come from the fact that Doppler surveys determine planetary masses while the transit searches measure planetary diameters. An interesting alternative to reconcile both estimations of 
planet occurrence has been proposed by \citet{Wolfgang:2011}: The apparent difference could be the result of (at least) two distinct types of planets resulting from multiple formation mechanisms involved in the production of super-Earths.

The observed planetary populations with their specific characteristics should help in constraining the physics of planetary formation. The discovery, by Doppler spectroscopy, of two planetary systems with super-Earths in the habitable zone (Gl\,581\,d and HD\,85512\,b) confirms the potential of that technique to identify planets around very close solar-type stars, which are suitable candidates for future experiments designed to detect spectroscopic signatures of life.

%----------------------------------------------------------------
\section{Acknowledgments} 
%----------------------------------------------------------------
We are grateful to the technical and scientific collaborators of the HARPS consortium at ESO who have contributed with their extraordinary passion and competences to the success of the HARPS project.  We acknowledge as well our colleagues of the Geneva Observatory for the dedicated work in the construction and maintenance of the EULER telescope and the CORALIE spectrograph during so many years. We would like to thank the Swiss National Foundation and the Geneva University for their continuous support. This research made use of the SIMBAD database, operated at the CDS, Strasbourg, France. NCS acknowledges the support by the European Research Council/European Community under the FP7 through Starting Grant agreement number 239953. NCS also acknowledges the support from Funda\c{c}\~ao para a Ci\^encia e a Tecnologia (FCT) through program Ci\^encia\,2007 funded by FCT/MCTES (Portugal) and POPH/FSE (EC), and in the form of grants reference PTDC/CTE-AST/098528/2008 and PTDC/CTE-AST/098604/2008. FB wish to thank the French National Research Agency (ANR-08-JCJC-0102-01).

%----------------------------------------------------------------
\section{References} 
%----------------------------------------------------------------
\bibliographystyle{aa} % style aa.bst
\bibliography{mayor.bib} % your references Yourfile.bib

%----------------------------------------------------------------

\longtab{3}{
\begin{longtable}{l c c c c l l}
\caption{Characteristics of the planetary systems orbiting stars in our sample.}\\
\hline\hline
Name & Period& Ecc& K& M2sin(i)& Reference & Special remarks\\
& [days]& & [$ms^{-1}$]& [Earth Mass] &  \\
\hline
\endfirsthead
\caption{continued.}\\
\hline\hline
Name & Period& Ecc& K& M2sin(i)& Reference & Special remarks\\
& [days]& & [$ms^{-1}$]& [Earth Mass]&  &  \\
\hline
\endhead
\hline
\multicolumn{7}{l}{[1] Excess residuals.}\\
\multicolumn{7}{l}{[2] Poor coverage or sampling inducing poorly constrained orbital parameters, in particular eccentricity.}\\
\multicolumn{7}{l}{[3] Peak in the RV GLS at periods corresponding to stellar rotation but not overlapping with orbital period of planetary}\\
\multicolumn{7}{l}{\hspace{5mm}companions. Most probably responsible for excess jitter.}\\
\multicolumn{7}{l}{[4] Peak at orbital period in the RV GLS has correspondence with peak at the same period in the R'HK GLS. However,}\\
\multicolumn{7}{l}{\hspace{5mm}after subtraction of the other companions, no correlation is detected between the RV residuals and R'HK.}\\
\multicolumn{7}{l}{[5] Semi-amplitude value of same level than (o-c) rms.}\\

\endfoot
\label{mayor_tab1}

HD142 b & 350.3 & 0.25 & 33.90 & 415.0 & Tinney2002 &     \\ 
HD1461 b & 5.77 & 0.00 & 2.44 & 6.94 & Rivera2010 &     \\ 
HD1461 c & 13.50 & 0.00 & 1.57 & 5.92 & This paper \& Segransan2011 &     \\ 
HD4113 b & 526.62 & 0.90 & 97.09 & 523.9 & Tamuz2008 &     \\ 
HD4208 b & 828.0 & 0.05 & 19.05 & 256.6 & Vogt2002 &     \\ 
HD4308 b & 15.62 & 0.08 & 4.00 & 13.45 & Udry2008 &     \\ 
HD6434 b & 21.99 & 0.17 & 34.20 & 126.2 & Mayor2004 &     \\ 
HD7199 b & 615 & 0.19 & 7.8 & 92 & Dumusque2011 &     \\ 
HD7449 b & 1275 & 0.82 & 41.6 & 353 & Dumusque2011 &     \\ 
HD7449 c & 4046 & 0.53 & 30 & 636 & Dumusque2011 &     \\ 
HD10180 b & 1.17 & 0.00 & 0.89 & 1.52 & Lovis2010 &     \\ 
HD10180 c & 5.75 & 0.07 & 4.53 & 13.2 & Lovis2010 &     \\ 
HD10180 d & 16.36 & 0.16 & 2.92 & 11.9 & Lovis2010 &     \\ 
HD10180 e & 49.75 & 0.06 & 4.16 & 24.8 & Lovis2010 &     \\ 
HD10180 f & 122.7 & 0.13 & 2.94 & 23.4 & Lovis2010 &     \\ 
HD10180 g & 595 & 0.00 & 1.62 & 22.1 & Lovis2010 &     \\ 
HD10180 h & 2149 & 0.15 & 3.21 & 67 & Lovis2010 &     \\ 
HD10647 b & 1003.0 & 0.15 & 17.89 & 294.0 & Butler2006 &     \\ 
HD11964A b & 37.94 & 0.21 & 3.56 & 16.8 & Wright2009 &     \\ 
HD11964A c & 2010 & 0.07 & 9.74 & 176 & Wright2009 &     \\ 
HD13808 b & 14.18 & 0.17 & 3.53 & 10.33 & This paper \& Queloz2011 &     \\ 
HD13808 c & 53.83 & 0.43 & 2.81 & 11.55 & This paper \& Queloz2011 &     \\ 
HD16141 b & 75.52 & 0.25 & 11.98 & 79.3 & Marcy2000 &     \\ 
HD16417 b & 16.43 & 0.22 & 3.33 & 14.7 & O'Toole2009 &     \\ 
HD19994 b & 466.2 & 0.26 & 29.29 & 421.7 & Mayor2004 &     \\ 
HD20003 b & 11.84 & 0.40 & 4.03 & 12.00 & This paper \& Udry2011 &     \\ 
HD20003 c & 33.82 & 0.16 & 2.95 & 13.42 & This paper \& Udry2011 &     \\ 
HD20781 b & 29.15 & 0.11 & 3.03 & 12.04 & This paper \& Udry2011 &     \\ 
HD20781 c & 85.13 & 0.28 & 2.88 & 15.78 & This paper \& Udry2011 &     \\ 
HD20782 b & 596.2 & 0.93 & 78.7 & 361 & Jones2006 &     \\ 
HD20794 b & 18.31 & 0.0 & 0.83 & 2.7 & Pepe2011 &     \\ 
HD20794 c & 40.11 & 0.0 & 0.56 & 2.4 & Pepe2011 &     \\ 
HD20794 d & 90.30 & 0.0 & 0.85 & 4.8 & Pepe2011 &     \\ 
HD21693 b & 22.65 & 0.26 & 2.73 & 10.22 & This paper \& Udry2011 &     \\ 
HD21693 c & 53.88 & 0.24 & 4.02 & 20.57 & This paper \& Udry2011 &     \\ 
HD23079 b & 730.6 & 0.10 & 54.90 & 776.6 & Tinney2002 &     \\ 
HD27631 b & 2220 & 0.17 & 27 & 540.3 & Marmier2011 &     \\ 
HD28185 b & 379.0 & 0.05 & 163.5 & 1842.5 & Santos2001 &     \\ 
HD30562 b & 1157.0 & 0.75 & 33.70 & 423.5 & Fischer2009 &     \\ 
HD31527 b & 16.54 & 0.13 & 3.01 & 11.55 & This paper \& Udry2011 &     \\ 
HD31527 c & 51.28 & 0.11 & 2.83 & 15.82 & This paper \& Udry2011 &     \\ 
HD31527 d & 274.49 & 0.38 & 1.79 & 16.50 & This paper \& Udry2011 &     \\ 
HD38858 b & 407.15 & 0.27 & 2.99 & 30.55 & This paper \& Benz2011 &     \\ 
HD39091 b & 2151.0 & 0.64 & 196.39 & 3206.3 & Jones2002 &     \\ 
HD39194 b & 5.63 & 0.20 & 1.95 & 3.71 & This paper \& Queloz2011 &     \\ 
HD39194 c & 14.02 & 0.11 & 2.26 & 5.94 & This paper \& Queloz2011 &     \\ 
HD39194 d & 33.94 & 0.20 & 1.49 & 5.13 & This paper \& Queloz2011 &   [5]  \\ 
HD40307 b & 4.31 & 0.19 & 1.94 & 4.07 & Mayor2009 &     \\ 
HD40307 c & 9.62 & 0.08 & 2.34 & 6.51 & Mayor2009 &     \\ 
HD40307 d & 20.43 & 0.11 & 2.48 & 8.85 & Mayor2009 &     \\ 
HD45184 b & 5.88 & 0.18 & 4.09 & 11.59 & This paper \& Udry2011 &     \\ 
HD45364 b & 226.0 & 0.18 & 7.44 & 61.1 & Correia2009 &     \\ 
HD45364 c & 344.3 & 0.12 & 22.97 & 219 & Correia2009 &     \\ 
HD47186 b & 4.08 & 0.04 & 9.28 & 23.0 & Bouchy2009 &     \\ 
HD47186 c & 3552 & 0.28 & 8.12 & 185 & Bouchy2009 &     \\ 
HD50499 b & 2457.87 & 0.25 & 23.02 & 554.5 & Vogt2005 &     \\ 
HD51608 b & 14.07 & 0.15 & 4.10 & 13.14 & This paper \& Udry2011 &     \\ 
HD51608 c & 95.41 & 0.41 & 3.25 & 17.97 & This paper \& Udry2011 &   [1],[3]  \\ 
HD52265 b & 119.29 & 0.32 & 42.09 & 340.5 & Butler2000 &     \\ 
HD65216 b & 579 & 0.26 & 37.8 & 449 & Mayor2004 &     \\ 
HD65216 c & 5542 & 0.15 & 27.5 & 712 & Marmier2011 &     \\ 
HD69830 b & 8.66 & 0.10 & 3.53 & 9.86 & Lovis2006 &     \\ 
HD69830 c & 31.58 & 0.10 & 2.77 & 11.98 & Lovis2006 &     \\ 
HD69830 d & 199.56 & 0.16 & 2.12 & 16.73 & Lovis2006 &     \\ 
HD70642 b & 2068.0 & 0.03 & 30.39 & 606.9 & Carter2003 &     \\ 
HD75289 b & 3.50 & 0.03 & 54.90 & 146.2 & Udry2000 &     \\ 
HD82943 c & 219.5 & 0.35 & 66.0 & 632.2 & Mayor2004 &     \\ 
HD82943 b & 441.2 & 0.21 & 43.59 & 551.0 & Mayor2004 &     \\ 
HD83443 b & 2.98 & 0.01 & 56.20 & 125.8 & Butler2002 &     \\ 
HD85390 b & 809.4 & 0.44 & 3.95 & 43.1 & Mordasini2010 &     \\ 
HD85512 b & 58.43 & 0.11 & 0.76 & 3.6 & Pepe2011 &     \\ 
HD86226 b & 1681 & 0.11 & 17.2 & 322 & Arriagada2010 &     \\ 
HD90156 b & 49.88 & 0.46 & 3.66 & 16.7 & Mordasini2010 &     \\ 
HD92788 b & 325.81 & 0.33 & 106.0 & 1132.7 & Fischer2001 &     \\ 
HD93083 b & 144.2 & 0.07 & 19.9 & 136.5 & Lovis2005 &     \\ 
HD93385 b & 13.18 & 0.15 & 2.21 & 8.36 & This paper \& Queloz2011 &     \\ 
HD93385 c & 46.02 & 0.24 & 1.82 & 10.12 & This paper \& Queloz2011 &     \\ 
HD96700 b & 8.12 & 0.10 & 3.02 & 9.02 & This paper \& Queloz2011 &   [1],[3]  \\ 
HD96700 c & 103.49 & 0.37 & 1.98 & 12.76 & This paper \& Queloz2011 &   [1],[3]  \\ 
HD98649 b & 10400 & 0.86 & 135 & 2225 & Marmier2011 &     \\ 
HD101930 b & 70.49 & 0.07 & 19.1 & 104.2 & Lovis2005 &     \\ 
HD102117 b & 20.82 & 0.07 & 11.1 & 48.0 & Lovis2005 &     \\ 
HD102365 b & 122.1 & 0.34 & 2.40 & 16.2 & Tinney2011 &     \\ 
HD104067 b & 55.83 & 0.06 & 10.9 & 51.4 & Segransan2011 &     \\ 
HD106515A b & 3630 & 0.60 & 174 & 3337 & Marmier2011 &     \\ 
HD108147 b & 10.89 & 0.52 & 25.10 & 82.0 & Pepe2002 &     \\ 
HD111232 b & 1118.0 & 0.18 & 162.0 & 2199.7 & Mayor2004 &     \\ 
HD114386 b & 445 & 0.12 & 11.9 & 119 & Mayor2004 &     \\ 
HD114386 c & 1046 & 0.06 & 28.1 & 377 & Marmier2011 &     \\ 
HD114729 b & 1114.0 & 0.16 & 18.79 & 300.3 & Butler2003 &     \\ 
HD114783 b & 493.7 & 0.14 & 31.89 & 351.2 & Vogt2002 &     \\ 
HD115617 b & 4.21 & 0.20 & 2.59 & 5.83 & Vogt2010 &     \\ 
HD115617 c & 38.07 & 0.18 & 4.68 & 22.20 & Vogt2010 &     \\ 
HD115617 d & 123.01 & 0.34 & 3.25 & 22.8 & Vogt2010 &     \\ 
HD117207 b & 2597.0 & 0.14 & 26.60 & 578.1 & Marcy2005 &     \\ 
HD117618 b & 25.82 & 0.41 & 12.80 & 56.1 & Tinney2005 &     \\ 
HD121504 b & 63.33 & 0.02 & 55.79 & 388.5 & Mayor2004 &     \\ 
HD126525 b & 948.12 & 0.13 & 5.11 & 71.33 & This paper \& Benz2011 &     \\ 
HD134060 b & 3.27 & 0.40 & 4.97 & 11.17 & This paper \& Udry2011 &     \\ 
HD134060 c & 1160.97 & 0.75 & 4.06 & 47.87 & This paper \& Udry2011 &     \\ 
HD134606 b & 12.08 & 0.15 & 2.68 & 9.27 & This paper \& Queloz2011 &     \\ 
HD134606 c & 59.51 & 0.29 & 2.17 & 12.14 & This paper \& Queloz2011 &   [4]  \\ 
HD134606 d & 459.26 & 0.46 & 3.66 & 38.52 & This paper \& Queloz2011 &     \\ 
HD134987 b & 258.18 & 0.23 & 49.5 & 496.9 & Vogt2000 &     \\ 
HD134987 c & 5000.0 & 0.11 & 9.30 & 255.8 & Vogt2000 &     \\ 
HD136352 b & 11.57 & 0.18 & 1.77 & 5.28 & This paper \& Udry2011 &  [5]   \\ 
HD136352 c & 27.58 & 0.16 & 2.82 & 11.38 & This paper \& Udry2011 &  [5]   \\ 
HD136352 d & 106.72 & 0.43 & 1.68 & 9.58 & This paper \& Udry2011 &  [5]   \\ 
HD137388 b & 330 & 0.36 & 7.9 & 71 & Dumusque2011 &     \\ 
HD141937 b & 653.21 & 0.40 & 234.5 & 3011.5 & Udry2002 &     \\ 
HD142022A b & 1928.0 & 0.52 & 92.0 & 1419.9 & Eggenberger2006 &     \\ 
HD147018 b & 44.23 & 0.46 & 145.33 & 676.1 & Segransan2010 &     \\ 
HD147018 c & 1008.0 & 0.13 & 141.19 & 2095.6 & Segransan2010 &     \\ 
HD150433 b & 1096.27 & -0.00 & 3.85 & 53.48 & This paper \& Benz2011 &  [2]   \\ 
HD154088 b & 18.59 & 0.38 & 1.78 & 6.14 & This paper \& Queloz2011 &     \\ 
HD156846 b & 359.51 & 0.84 & 464.29 & 3498.5 & Tamuz2008 &     \\ 
HD157172 b & 105.0 & 0.33 & 5.46 & 34.6 & This paper \& Benz2011 &     \\ 
HD160691 b & 9.64 & 0.12 & 3.05 & 10.6 & Pepe2007 &     \\ 
HD160691 c & 313.2 & 0.04 & 16.9 & 189.1 & Pepe2007 &     \\ 
HD160691 d & 648.7 & 0.18 & 39.0 & 546.9 & Pepe2007 &     \\ 
HD160691 e & 8723 & 0.43 & 25.8 & 790 & Pepe2007 &     \\ 
HD166724 b & 8100 & 0.77 & 72 & 1310 & Marmier2011 &     \\ 
HD168443 b & 58.11 & 0.52 & 475.53 & 2465.3 & Marcy1999 &     \\ 
HD168443 c & 1748.15 & 0.21 & 298.13 & 5571.5 & Marcy2001 &     \\ 
HD168746 b & 6.40 & 0.10 & 28.60 & 77.9 & Pepe2002 &     \\ 
HD169830 b & 225.62 & 0.31 & 80.69 & 918.4 & Naef2001 &     \\ 
HD169830 c & 2100.0 & 0.33 & 54.29 & 1291.3 & Mayor2004 &     \\ 
HD179949 b & 3.09 & 0.02 & 112.59 & 286.7 & Tinney2001 &     \\ 
HD181433 b & 9.37 & 0.42 & 2.89 & 7.32 & Bouchy2009 &     \\ 
HD181433 c & 1019 & 0.25 & 17.2 & 222 & Bouchy2009 &     \\ 
HD181433 d & 3201 & 0.11 & 9.49 & 184 & Bouchy2009 &     \\ 
HD187085 b & 986.0 & 0.46 & 17.0 & 255.4 & Jones2006 &     \\ 
HD189567 b & 14.27 & 0.23 & 3.02 & 10.03 & This paper \& Queloz2011 &     \\ 
HD192310 b & 74.72 & 0.13 & 3.00 & 16.9 & Howard2010 &     \\ 
HD192310 c & 525.8 & 0.32 & 2.27 & 24 & Pepe2011 &     \\ 
HD196050 b & 1378.0 & 0.22 & 49.70 & 903.7 & Jones2002 &     \\ 
HD196067 b & 4100 & 0.63 & 112 & 2257 & Marmier2011 &     \\ 
HD204313 c & 34.88 & 0.13 & 3.28 & 16.9 & This paper \& Segransan2011 &   [4]  \\ 
HD204313 b & 2132 & 0.11 & 72.9 & 1479 & Segransan2009 &     \\ 
HD204941 b & 1733 & 0.37 & 5.94 & 85 & Dumusque2011 &     \\ 
HD208487 b & 130.08 & 0.23 & 19.70 & 162.8 & Tinney2005 &     \\ 
HD210277 b & 442.19 & 0.47 & 38.93 & 404.5 & Marcy1999 &     \\ 
HD213240 b & 882.7 & 0.42 & 96.59 & 1440.6 & Santos2001 &     \\ 
HD215152 b & 7.28 & 0.34 & 1.26 & 2.77 & This paper \& Queloz2011 &   [5]  \\ 
HD215152 c & 10.86 & 0.38 & 1.26 & 3.09 & This paper \& Queloz2011 &   [5]  \\ 
HD215456 b & 192.0 & 0.13 & 3.62 & 32.6 & This paper \& Benz2011 &     \\ 
HD215456 c & 2268 & 0.17 & 3.77 & 76.8 & This paper \& Benz2011 &     \\ 
HD216435 b & 1311.0 & 0.07 & 19.60 & 386.1 & Jones2003 &     \\ 
HD216437 b & 1353.0 & 0.31 & 39.0 & 689.1 & Jones2002 &     \\ 
HD216770 b & 118.45 & 0.37 & 30.89 & 205.6 & Mayor2004 &     \\ 
HD217107 b & 7.12 & 0.12 & 139.20 & 445.3 & Fischer1999 &     \\ 
HD217107 c & 4270.0 & 0.51 & 35.70 & 831.3 & Vogt2005 &     \\ 
HD218566 b & 225.7 & 0.30 & 8.30 & 67.7 & Meschiari2011 &     \\ 
HD220689 b & 2191 & 0.2 & 19. & 377 & Marmier2011 &     \\ 
HD222582 b & 572.38 & 0.72 & 276.29 & 2425.1 & Vogt2000 &     \\ 

\end{longtable}
}% End \longtab

%---------------------------  HD1461  ----------
 
 \begin{table*}
 \caption{Orbital and physical parameters of the planets orbiting HD\,1461 as obtained from a Keplerian fit to the data. }
 %\label{table_hd1461_par}
 \begin{center}
 \begin{tabular}{l l c }
 \hline \hline
 \noalign{\smallskip}
 {\bf Parameter}	& {\bf [unit]}		& {\bf HD 1461 c}	 \\
 \hline 
 \noalign{\smallskip}
 %Epoch		& [BJD]			& \multicolumn{1}{c}{2'454'783.40362208}    \\ 
 %$i$			& [deg]			& \multicolumn{1}{c}{$ 90 $ (fixed) }  \\  
 $V$			& [km\,s$^{-1}$]		& \multicolumn{1}{c}{$ -10.0670\,(\pm 0.0002) $}  \\
 \hline 
 \noalign{\smallskip}
 $P$			& [days]			& $13.505$		 \\ 
 			&				& $(\pm 0.004)$	 \\ 
 $e$			&				& $ 0.0 $			\\ 
 			&				& $(fixed)$		 \\ 
 $K$			& [m\,s$^{-1}$]		& $ 1.57 $		 \\  
 			&				& $(\pm 0.19)   $	 \\
 \hline
 \noalign{\smallskip}
 $m \sin i$		& [$M_\oplus$]		& $ 5.92$		 \\
 			&				& $(\pm 0.76) $	 \\
 $a$			& [AU]			& $ 0.1117 $		 \\
 			&				& $(\pm 0.0018)   $	 \\
 \hline
 \noalign{\smallskip}
 $N_\mathrm{meas}$ &			& \multicolumn{1}{c}{167}  \\
 Span		& [days]			& \multicolumn{1}{c}{2856} \\
 rms			& [m\,s$^{-1}$]		& \multicolumn{1}{c}{1.70} \\
 $\chi_r^2$	&				& \multicolumn{1}{c}{3.61} \\
 \hline
 \end{tabular}
 \end{center}
 \end{table*}
 
 %----------------------------------------------------------------------
 
 %---------------------------  HD13808  ----------
 
 \begin{table*}
 \caption{Orbital and physical parameters of the planets orbiting HD\,13808 as obtained from a Keplerian fit to the data. }
 %\label{table_hd13808_par}
 \begin{center}
 \begin{tabular}{l l c c }
 \hline \hline
 \noalign{\smallskip}
 {\bf Parameter}	& {\bf [unit]}		& {\bf HD 13808 b}	& {\bf HD 13808 c}	 \\
 \hline 
 \noalign{\smallskip}
 %Epoch		& [BJD]			& \multicolumn{2}{c}{2'454'783.40362208}    \\ 
 %$i$			& [deg]			& \multicolumn{2}{c}{$ 90 $ (fixed) }  \\  
 $V$			& [km\,s$^{-1}$]		& \multicolumn{2}{c}{$ 41.0949\,(\pm 0.0005) $}  \\
 \hline 
 \noalign{\smallskip}
 $P$			& [days]			& $14.182$		& $53.83$		 \\ 
 			&				& $(\pm 0.005)$	& $(\pm 0.11)$	 \\ 
 $e$			&				& $ 0.17 $			& $ 0.43 $			\\ 
 			&				& $(\pm 0.07)$		& $ (\pm 0.20)$		 \\ 
 $K$			& [m\,s$^{-1}$]		& $ 3.53 $		& $ 2.81 $		 \\  
 			&				& $(\pm 0.29)   $	& $(\pm 0.46)  $	 \\
 \hline
 \noalign{\smallskip}
 $m \sin i$		& [$M_\oplus$]		& $ 10.33$		& $ 11.55 $		 \\
 			&				& $(\pm 0.92) $		& $(\pm 1.62) $	 \\
 $a$			& [AU]			& $ 0.1017 $		& $ 0.2476 $		 \\
 			&				& $(\pm 0.0016)   $	& $(\pm 0.0041)$	 \\
 \hline
 \noalign{\smallskip}
 $N_\mathrm{meas}$ &			& \multicolumn{2}{c}{133}  \\
 Span		& [days]			& \multicolumn{2}{c}{2770} \\
 rms			& [m\,s$^{-1}$]		& \multicolumn{2}{c}{1.93} \\
 $\chi_r^2$	&				& \multicolumn{2}{c}{4.10} \\
 \hline
 \end{tabular}
 \end{center}
 \end{table*}
 
 %----------------------------------------------------------------------
 
 %---------------------------  HD20003  ----------
 
 \begin{table*}
 \caption{Orbital and physical parameters of the planets orbiting HD\,20003 as obtained from a Keplerian fit to the data. }
 %\label{table_hd20003_par}
 \begin{center}
 \begin{tabular}{l l c c }
 \hline \hline
 \noalign{\smallskip}
 {\bf Parameter}	& {\bf [unit]}		& {\bf HD 20003 b}	& {\bf HD 20003 c}	 \\
 \hline 
 \noalign{\smallskip}
 %Epoch		& [BJD]			& \multicolumn{2}{c}{2'454'783.40362208}    \\ 
 %$i$			& [deg]			& \multicolumn{2}{c}{$ 90 $ (fixed) }  \\  
 $V$			& [km\,s$^{-1}$]		& \multicolumn{2}{c}{$ -16.1040\,(\pm 0.0006) $}  \\
 \hline 
 \noalign{\smallskip}
 $P$			& [days]			& $11.849$		& $33.823$		 \\ 
 			&				& $(\pm 0.003)$	& $(\pm 0.065)$	 \\ 
 $e$			&				& $ 0.40 $			& $ 0.16 $			\\ 
 			&				& $(\pm 0.08)$		& $ (\pm 0.09)$		 \\ 
 $K$			& [m\,s$^{-1}$]		& $ 4.03 $		& $ 2.95 $		 \\  
 			&				& $(\pm 0.33)   $	& $(\pm 0.28)  $	 \\
 \hline
 \noalign{\smallskip}
 $m \sin i$		& [$M_\oplus$]		& $ 12.00$		& $ 13.42 $		 \\
 			&				& $(\pm 0.97) $	& $(\pm 1.28) $	 \\
 $a$			& [AU]			& $ 0.0974 $		& $ 0.1961 $		 \\
 			&				& $(\pm 0.0016)   $	& $(\pm 0.0032)$	 \\
 \hline
 \noalign{\smallskip}
 $N_\mathrm{meas}$ &			& \multicolumn{2}{c}{104}  \\
 Span		& [days]			& \multicolumn{2}{c}{2770} \\
 rms			& [m\,s$^{-1}$]		& \multicolumn{2}{c}{1.56} \\
 $\chi_r^2$	&				& \multicolumn{2}{c}{2.59} \\
 \hline
 \end{tabular}
 \end{center}
 \end{table*}
 
 %----------------------------------------------------------------------
 
 %---------------------------  HD20781  ----------
 
 \begin{table*}
 \caption{Orbital and physical parameters of the planets orbiting HD\,20781 as obtained from a Keplerian fit to the data. }
 %\label{table_hd20781_par}
 \begin{center}
 \begin{tabular}{l l c c }
 \hline \hline
 \noalign{\smallskip}
 {\bf Parameter}	& {\bf [unit]}		& {\bf HD 20781 b}	& {\bf HD 20781 c}	 \\
 \hline 
 \noalign{\smallskip}
 %Epoch		& [BJD]			& \multicolumn{2}{c}{2'454'783.40362208}    \\ 
 %$i$			& [deg]			& \multicolumn{2}{c}{$ 90 $ (fixed) }  \\  
 $V$			& [km\,s$^{-1}$]		& \multicolumn{2}{c}{$ 40.3671\,(\pm 0.0005) $}  \\
 \hline 
 \noalign{\smallskip}
 $P$			& [days]			& $29.15$			& $85.13$		 \\ 
 			&				& $(\pm 0.02)$		& $(\pm 0.12)$	 \\ 
 $e$			&				& $ 0.11 $			& $ 0.28 $			\\ 
 			&				& $(\pm 0.06)$		& $ (\pm 0.09)$		 \\ 
 $K$			& [m\,s$^{-1}$]		& $ 3.03 $		& $ 2.88 $		 \\  
 			&				& $(\pm 0.26)   $	& $(\pm 0.23)  $	 \\
 \hline
 \noalign{\smallskip}
 $m \sin i$		& [$M_\oplus$]		& $ 12.04$		& $ 15.78 $		 \\
 			&				& $(\pm 1.12)	 $	& $(\pm 1.21) $	 \\
 $a$			& [AU]			& $ 0.1690 $		& $ 0.3456 $		 \\
 			&				& $(\pm 0.0028)   $	& $(\pm 0.0057)$	 \\
 \hline
 \noalign{\smallskip}
 $N_\mathrm{meas}$ &			& \multicolumn{2}{c}{96}  \\
 Span		& [days]			& \multicolumn{2}{c}{2647} \\
 rms			& [m\,s$^{-1}$]		& \multicolumn{2}{c}{1.12} \\
 $\chi_r^2$	&				& \multicolumn{2}{c}{1.39} \\
 \hline
 \end{tabular}
 \end{center}
 \end{table*}
 
 %----------------------------------------------------------------------
 
 %---------------------------  HD21693  ----------
 
 \begin{table*}
 \caption{Orbital and physical parameters of the planets orbiting HD\,21693 as obtained from a Keplerian fit to the data. }
 %\label{table_hd21693_par}
 \begin{center}
 \begin{tabular}{l l c c }
 \hline \hline
 \noalign{\smallskip}
 {\bf Parameter}	& {\bf [unit]}		& {\bf HD 21693 b}	& {\bf HD 21693 c}	 \\
 \hline 
 \noalign{\smallskip}
 %Epoch		& [BJD]			& \multicolumn{2}{c}{2'454'783.40362208}    \\ 
 %$i$			& [deg]			& \multicolumn{2}{c}{$ 90 $ (fixed) }  \\  
 $V$			& [km\,s$^{-1}$]		& \multicolumn{2}{c}{$ 39.7685\,(\pm 0.0008) $}  \\
 \hline 
 \noalign{\smallskip}
 $P$			& [days]			& $22.656$		& $53.88$		 \\ 
 			&				& $(\pm 0.024)$	& $(\pm 0.07)$	 \\ 
 $e$			&				& $ 0.26 $			& $ 0.24 $			\\ 
 			&				& $(\pm 0.17)$		& $ (\pm 0.09)$		 \\ 
 $K$			& [m\,s$^{-1}$]		& $ 2.73 $		& $ 4.02 $		 \\  
 			&				& $(\pm 0.33)   $	& $(\pm 0.35)  $	 \\
 \hline
 \noalign{\smallskip}
 $m \sin i$		& [$M_\oplus$]		& $ 10.22$		& $ 20.57 $		 \\
 			&				& $(\pm 1.46) $		& $(\pm 1.80) $	 \\
 $a$			& [AU]			& $ 0.1484 $		& $ 0.2644 $		 \\
 			&				& $(\pm 0.0024)   $	& $(\pm 0.0044)$	 \\
 \hline
 \noalign{\smallskip}
 $N_\mathrm{meas}$ &			& \multicolumn{2}{c}{128}  \\
 Span		& [days]			& \multicolumn{2}{c}{2813} \\
 rms			& [m\,s$^{-1}$]		& \multicolumn{2}{c}{2.02} \\
 $\chi_r^2$	&				& \multicolumn{2}{c}{5.16} \\
 \hline
 \end{tabular}
 \end{center}
 \end{table*}
 
 %----------------------------------------------------------------------
 
 %---------------------------  HD31527  ----------
 
 \begin{table*}
 \caption{Orbital and physical parameters of the planets orbiting HD\,31527 as obtained from a Keplerian fit to the data. }
 %\label{table_hd31527_par}
 \begin{center}
 \begin{tabular}{l l c c c}
 \hline \hline
 \noalign{\smallskip}
 {\bf Parameter}	& {\bf [unit]}		& {\bf HD 31527 b}	& {\bf HD 31527 c}	& {\bf HD 31527 d} \\
 \hline 
 \noalign{\smallskip}
 %Epoch		& [BJD]			& \multicolumn{3}{c}{2'454'783.40362208}    \\ 
 %$i$			& [deg]			& \multicolumn{3}{c}{$ 90 $ (fixed) }  \\  
 $V$			& [km\,s$^{-1}$]		& \multicolumn{3}{c}{$ 25.7391\,(\pm 0.0004) $}  \\
 \hline 
 \noalign{\smallskip}
 $P$			& [days]			& $16.546$		& $51.28$			& $274.5$		 \\ 
 			&				& $(\pm 0.007)$	& $(\pm 0.09)$		& $(\pm 7.8)$	 \\ 
 $e$			&				& $ 0.13 $			& $ 0.11 $			& $ 0.38 $		 \\ 
 			&				& $(\pm 0.05)$		& $ (\pm 0.07)$		& $(\pm 0.25)$	 \\ 
 $K$			& [m\,s$^{-1}$]		& $ 3.01 $		& $ 2.83 $		& $ 1.79 $		 \\  
 			&				& $(\pm 0.18)   $	& $(\pm 0.17)  $	& $(\pm 0.68)  $	 \\
 \hline
 \noalign{\smallskip}
 $m \sin i$		& [$M_\oplus$]		& $ 11.55$		& $ 15.82 $		& $ 16.50 $		 \\
 			&				& $(\pm 0.79) $		& $(\pm 1.10) $		& $(\pm 3.04)  $	 \\
 $a$			& [AU]			& $ 0.1253 $		& $ 0.2665 $		& $ 0.818 $		 \\
 			&				& $(\pm 0.0020)   $	& $(\pm 0.0044)$	& $(\pm 0.020)  $	 \\
 \hline
 \noalign{\smallskip}
 $N_\mathrm{meas}$ &			& \multicolumn{3}{c}{167}  \\
 Span		& [days]			& \multicolumn{3}{c}{2719} \\
 rms			& [m\,s$^{-1}$]		& \multicolumn{3}{c}{1.35} \\
 $\chi_r^2$	&				& \multicolumn{3}{c}{2.14} \\
 \hline
 \end{tabular}
 \end{center}
 \end{table*}
 
 %----------------------------------------------------------------------
 
 %---------------------------  HD38858  ----------
 
 \begin{table*}
 \caption{Orbital and physical parameters of the planets orbiting HD\,38858 as obtained from a Keplerian fit to the data. }
 %\label{table_hd38858_par}
 \begin{center}
 \begin{tabular}{l l c }
 \hline \hline
 \noalign{\smallskip}
 {\bf Parameter}	& {\bf [unit]}		& {\bf HD 38858 b}	 \\
 \hline 
 \noalign{\smallskip}
 %Epoch		& [BJD]			& \multicolumn{1}{c}{2'454'783.40362208}    \\ 
 %$i$			& [deg]			& \multicolumn{1}{c}{$ 90 $ (fixed) }  \\  
 $V$			& [km\,s$^{-1}$]		& \multicolumn{1}{c}{$ 31.6383\,(\pm 0.0004) $}  \\
 \hline 
 \noalign{\smallskip}
 $P$			& [days]			& $407.1$		 \\ 
 			&				& $(\pm 4.3)$	 \\ 
 $e$			&				& $ 0.27 $			\\ 
 			&				& $(\pm 0.17)$		 \\ 
 $K$			& [m\,s$^{-1}$]		& $ 2.99 $		 \\  
 			&				& $(\pm 0.33)   $	 \\
 \hline
 \noalign{\smallskip}
 $m \sin i$		& [$M_\oplus$]		& $ 30.55$		 \\
 			&				& $(\pm 4.11) $	 \\
 $a$			& [AU]			& $ 1.038 $		 \\
 			&				& $(\pm 0.019)   $	 \\
 \hline
 \noalign{\smallskip}
 $N_\mathrm{meas}$ &			& \multicolumn{1}{c}{52}  \\
 Span		& [days]			& \multicolumn{1}{c}{2964} \\
 rms			& [m\,s$^{-1}$]		& \multicolumn{1}{c}{1.38} \\
 $\chi_r^2$	&				& \multicolumn{1}{c}{2.53} \\
 \hline
 \end{tabular}
 \end{center}
 \end{table*}
 
 %----------------------------------------------------------------------
 
 %---------------------------  HD39194  ----------
 
 \begin{table*}
 \caption{Orbital and physical parameters of the planets orbiting HD\,39194 as obtained from a Keplerian fit to the data. }
 %\label{table_hd39194_par}
 \begin{center}
 \begin{tabular}{l l c c c}
 \hline \hline
 \noalign{\smallskip}
 {\bf Parameter}	& {\bf [unit]}		& {\bf HD 39194 b}	& {\bf HD 39194 c}	& {\bf HD 39194 d} \\
 \hline 
 \noalign{\smallskip}
 %Epoch		& [BJD]			& \multicolumn{3}{c}{2'454'783.40362208}    \\ 
 %$i$			& [deg]			& \multicolumn{3}{c}{$ 90 $ (fixed) }  \\  
 $V$			& [km\,s$^{-1}$]		& \multicolumn{3}{c}{$ 14.1692\,(\pm 0.0005) $}  \\
 \hline 
 \noalign{\smallskip}
 $P$			& [days]			& $5.6363$		& $14.025$		& $33.941$		 \\ 
 			&				& $(\pm 0.0008)$	& $(\pm 0.005)$	& $(\pm 0.035)$	 \\ 
 $e$			&				& $ 0.20 $			& $ 0.11 $			& $ 0.20 $		 \\ 
 			&				& $(\pm 0.10)$		& $ (\pm 0.06)$		& $(\pm 0.16)$	 \\ 
 $K$			& [m\,s$^{-1}$]		& $ 1.95 $		& $ 2.26 $		& $ 1.49 $		 \\  
 			&				& $(\pm 0.16)   $	& $(\pm 0.15)  $	& $(\pm 0.17)  $	 \\
 \hline
 \noalign{\smallskip}
 $m \sin i$		& [$M_\oplus$]		& $ 3.72$			& $ 5.94 $			& $ 5.14 $		 \\
 			&				& $(\pm 0.33) $		& $(\pm 0.47) $		& $(\pm 0.66)  $	 \\
 $a$			& [AU]			& $ 0.0519 $		& $ 0.0954 $		& $ 0.1720 $		 \\
 			&				& $(\pm 0.0008)   $	& $(\pm 0.0016)$	& $(\pm 0.0029)  $	 \\
 \hline
 \noalign{\smallskip}
 $N_\mathrm{meas}$ &			& \multicolumn{3}{c}{133}  \\
 Span		& [days]			& \multicolumn{3}{c}{2717} \\
 rms			& [m\,s$^{-1}$]		& \multicolumn{3}{c}{1.11} \\
 $\chi_r^2$	&				& \multicolumn{3}{c}{1.30} \\
 \hline
 \end{tabular}
 \end{center}
 \end{table*}
 
 %----------------------------------------------------------------------
 
 %---------------------------  HD45184  ----------
 
 \begin{table*}
 \caption{Orbital and physical parameters of the planets orbiting HD\,45184 as obtained from a Keplerian fit to the data. }
 %\label{table_hd45184_par}
 \begin{center}
 \begin{tabular}{l l c }
 \hline \hline
 \noalign{\smallskip}
 {\bf Parameter}	& {\bf [unit]}		& {\bf HD 45184 b}	 \\
 \hline 
 \noalign{\smallskip}
 %Epoch		& [BJD]			& \multicolumn{1}{c}{2'454'783.40362208}    \\ 
 %$i$			& [deg]			& \multicolumn{1}{c}{$ 90 $ (fixed) }  \\  
 $V$			& [km\,s$^{-1}$]		& \multicolumn{1}{c}{$ -3.7584\,(\pm 0.0008) $}  \\
 \hline 
 \noalign{\smallskip}
 $P$			& [days]			& $5.8872$		 \\ 
 			&				& $(\pm 0.0015)$	 \\ 
 $e$			&				& $ 0.30 $			\\ 
 			&				& $(\pm 0.19)$		 \\ 
 $K$			& [m\,s$^{-1}$]		& $ 4.77 $		 \\  
 			&				& $(\pm 1.18)   $	 \\
 \hline
 \noalign{\smallskip}
 $m \sin i$		& [$M_\oplus$]		& $ 12.73$		 \\
 			&				& $(\pm 1.67) $	 \\
 $a$			& [AU]			& $ 0.0638$		 \\
 			&				& $(\pm 0.0010)   $	 \\
 \hline
 \noalign{\smallskip}
 $N_\mathrm{meas}$ &			& \multicolumn{1}{c}{82}  \\
 Span		& [days]			& \multicolumn{1}{c}{2738} \\
 rms			& [m\,s$^{-1}$]		& \multicolumn{1}{c}{3.81} \\
 $\chi_r^2$	&				& \multicolumn{1}{c}{21.86} \\
 \hline
 \end{tabular}
 \end{center}
 \end{table*}
 
 %----------------------------------------------------------------------
 
 %---------------------------  HD51608  ----------
 
 \begin{table*}
 \caption{Orbital and physical parameters of the planets orbiting HD\,51608 as obtained from a Keplerian fit to the data. }
 %\label{table_hd51608_par}
 \begin{center}
 \begin{tabular}{l l c c }
 \hline \hline
 \noalign{\smallskip}
 {\bf Parameter}	& {\bf [unit]}		& {\bf HD 51608 b}	& {\bf HD 51608 c}	 \\
 \hline 
 \noalign{\smallskip}
 %Epoch		& [BJD]			& \multicolumn{2}{c}{2'454'783.40362208}    \\ 
 %$i$			& [deg]			& \multicolumn{2}{c}{$ 90 $ (fixed) }  \\  
 $V$			& [km\,s$^{-1}$]		& \multicolumn{2}{c}{$ 39.9773\,(\pm 0.0008) $}  \\
 \hline 
 \noalign{\smallskip}
 $P$			& [days]			& $14.070$		& $95.42$		 \\ 
 			&				& $(\pm 0.004)$	& $(\pm 0.39)$	 \\ 
 $e$			&				& $ 0.15 $			& $ 0.41 $			\\ 
 			&				& $(\pm 0.06)$		& $ (\pm 0.18)$		 \\ 
 $K$			& [m\,s$^{-1}$]		& $ 4.10 $		& $ 3.25 $		 \\  
 			&				& $(\pm 0.27)   $	& $(\pm 0.61)  $	 \\
 \hline
 \noalign{\smallskip}
 $m \sin i$		& [$M_\oplus$]		& $ 13.14$		& $ 17.97 $		 \\
 			&				& $(\pm 0.98) $	& $(\pm 2.61) $	 \\
 $a$			& [AU]			& $ 0.1059 $		& $ 0.379 $		 \\
 			&				& $(\pm 0.0017)   $	& $(\pm 0.006)$	 \\
 \hline
 \noalign{\smallskip}
 $N_\mathrm{meas}$ &			& \multicolumn{2}{c}{118}  \\
 Span		& [days]			& \multicolumn{2}{c}{2697} \\
 rms			& [m\,s$^{-1}$]		& \multicolumn{2}{c}{1.82} \\
 $\chi_r^2$	&				& \multicolumn{2}{c}{3.95} \\
 \hline
 \end{tabular}
 \end{center}
 \end{table*}
 
 %----------------------------------------------------------------------
 
 %---------------------------  HD93385  ----------
 
 \begin{table*}
 \caption{Orbital and physical parameters of the planets orbiting HD\,93385 as obtained from a Keplerian fit to the data. }
 %\label{table_hd93385_par}
 \begin{center}
 \begin{tabular}{l l c c }
 \hline \hline
 \noalign{\smallskip}
 {\bf Parameter}	& {\bf [unit]}		& {\bf HD 93385 b}	& {\bf HD 93385 c}	 \\
 \hline 
 \noalign{\smallskip}
 %Epoch		& [BJD]			& \multicolumn{2}{c}{2'454'783.40362208}    \\ 
 %$i$			& [deg]			& \multicolumn{2}{c}{$ 90 $ (fixed) }  \\  
 $V$			& [km\,s$^{-1}$]		& \multicolumn{2}{c}{$ 47.5791\,(\pm 0.0005) $}  \\
 \hline 
 \noalign{\smallskip}
 $P$			& [days]			& $13.186$		& $46.025$		 \\ 
 			&				& $(\pm 0.006)$	& $(\pm 0.073)$	 \\ 
 $e$			&				& $ 0.15 $			& $ 0.24 $			\\ 
 			&				& $(\pm 0.11)$		& $ (\pm 0.18)$		 \\ 
 $K$			& [m\,s$^{-1}$]		& $ 2.21 $		& $ 1.82 $		 \\  
 			&				& $(\pm 0.23)   $	& $(\pm 0.96)  $	 \\
 \hline
 \noalign{\smallskip}
 $m \sin i$		& [$M_\oplus$]		& $ 8.36$		& $ 10.12 $		 \\
 			&				& $(\pm 0.88) $	& $(\pm 1.47) $	 \\
 $a$			& [AU]			& $ 0.1116 $		& $ 0.2570 $		 \\
 			&				& $(\pm 0.0018)   $	& $(\pm 0.0043)$	 \\
 \hline
 \noalign{\smallskip}
 $N_\mathrm{meas}$ &			& \multicolumn{2}{c}{127}  \\
 Span		& [days]			& \multicolumn{2}{c}{2733} \\
 rms			& [m\,s$^{-1}$]		& \multicolumn{2}{c}{1.42} \\
 $\chi_r^2$	&				& \multicolumn{2}{c}{2.38} \\
 \hline
 \end{tabular}
 \end{center}
 \end{table*}
 
 %----------------------------------------------------------------------
 
 %---------------------------  HD96700  ----------
 
 \begin{table*}
 \caption{Orbital and physical parameters of the planets orbiting HD\,96700 as obtained from a Keplerian fit to the data. }
 %\label{table_hd96700_par}
 \begin{center}
 \begin{tabular}{l l c c }
 \hline \hline
 \noalign{\smallskip}
 {\bf Parameter}	& {\bf [unit]}		& {\bf HD 96700 b}	& {\bf HD 96700 c}	 \\
 \hline 
 \noalign{\smallskip}
 %Epoch		& [BJD]			& \multicolumn{2}{c}{2'454'783.40362208}    \\ 
 %$i$			& [deg]			& \multicolumn{2}{c}{$ 90 $ (fixed) }  \\  
 $V$			& [km\,s$^{-1}$]		& \multicolumn{2}{c}{$ 12.8584\,(\pm 0.0071) $}  \\
 \hline 
 \noalign{\smallskip}
 $P$			& [days]			& $8.1256$		& $103.49$		 \\ 
 			&				& $(\pm 0.0013)$	& $(\pm 0.58)$	 \\ 
 $e$			&				& $ 0.10 $			& $ 0.37 $			\\ 
 			&				& $(\pm 0.05)$		& $ (\pm 0.19)$		 \\ 
 $K$			& [m\,s$^{-1}$]		& $ 3.02 $		& $ 1.98 $		 \\  
 			&				& $(\pm 0.18)   $	& $(\pm 0.37)  $	 \\
 \hline
 \noalign{\smallskip}
 $m \sin i$		& [$M_\oplus$]		& $ 9.03$		& $ 12.76 $		 \\
 			&				& $(\pm 0.63) $	& $(\pm 1.63) $	 \\
 $a$			& [AU]			& $ 0.0774 $		& $ 0.422 $		 \\
 			&				& $(\pm 0.0012)   $	& $(\pm 0.007)$	 \\
 \hline
 \noalign{\smallskip}
 $N_\mathrm{meas}$ &			& \multicolumn{2}{c}{146}  \\
 Span		& [days]			& \multicolumn{2}{c}{2717} \\
 rms			& [m\,s$^{-1}$]		& \multicolumn{2}{c}{1.48} \\
 $\chi_r^2$	&				& \multicolumn{2}{c}{2.93} \\
 \hline
 \end{tabular}
 \end{center}
 \end{table*}
 
 %----------------------------------------------------------------------
 
 %---------------------------  HD126525  ----------
 
 \begin{table*}
 \caption{Orbital and physical parameters of the planets orbiting HD\,126525 as obtained from a Keplerian fit to the data. }
 %\label{table_hd126525_par}
 \begin{center}
 \begin{tabular}{l l c }
 \hline \hline
 \noalign{\smallskip}
 {\bf Parameter}	& {\bf [unit]}		& {\bf HD 126525 b}	 \\
 \hline 
 \noalign{\smallskip}
 %Epoch		& [BJD]			& \multicolumn{1}{c}{2'454'783.40362208}    \\ 
 %$i$			& [deg]			& \multicolumn{1}{c}{$ 90 $ (fixed) }  \\  
 $V$			& [km\,s$^{-1}$]		& \multicolumn{1}{c}{$ 13.1384\,(\pm 0.0006) $}  \\
 \hline 
 \noalign{\smallskip}
 $P$			& [days]			& $948.1$		 \\ 
 			&				& $(\pm 22.0)$	 \\ 
 $e$			&				& $ 0.13 $			\\ 
 			&				& $(\pm 0.07)$		 \\ 
 $K$			& [m\,s$^{-1}$]		& $ 5.11 $		 \\  
 			&				& $(\pm 0.34)   $	 \\
 \hline
 \noalign{\smallskip}
 $m \sin i$		& [$M_\oplus$]		& $ 71.33$		 \\
 			&				& $(\pm 5.78) $	 \\
 $a$			& [AU]			& $ 1.811 $		 \\
 			&				& $(\pm 0.041)   $	 \\
 \hline
 \noalign{\smallskip}
 $N_\mathrm{meas}$ &			& \multicolumn{1}{c}{45}  \\
 Span		& [days]			& \multicolumn{1}{c}{2575} \\
 rms			& [m\,s$^{-1}$]		& \multicolumn{1}{c}{1.27} \\
 $\chi_r^2$	&				& \multicolumn{1}{c}{2.10} \\
 \hline
 \end{tabular}
 \end{center}
 \end{table*}
 
 %----------------------------------------------------------------------
 
 %---------------------------  HD134060  ----------
 
 \begin{table*}
 \caption{Orbital and physical parameters of the planets orbiting HD\,134060 as obtained from a Keplerian fit to the data. }
 %\label{table_hd134060_par}
 \begin{center}
 \begin{tabular}{l l c c }
 \hline \hline
 \noalign{\smallskip}
 {\bf Parameter}	& {\bf [unit]}		& {\bf HD 134060 b}	& {\bf HD 134060 c}	 \\
 \hline 
 \noalign{\smallskip}
 %Epoch		& [BJD]			& \multicolumn{2}{c}{2'454'783.40362208}    \\ 
 %$i$			& [deg]			& \multicolumn{2}{c}{$ 90 $ (fixed) }  \\  
 $V$			& [km\,s$^{-1}$]		& \multicolumn{2}{c}{$ 37.9893\,(\pm 0.0026) $}  \\
 \hline 
 \noalign{\smallskip}
 $P$			& [days]			& $3.2700$		& $1161$		 \\ 
 			&				& $(\pm 0.0002)$	& $(\pm 27)$	 \\ 
 $e$			&				& $ 0.40 $			& $ 0.75 $			\\ 
 			&				& $(\pm 0.04)$		& $ (\pm 0.19)$		 \\ 
 $K$			& [m\,s$^{-1}$]		& $ 4.97 $		& $ 4.1 $		 \\  
 			&				& $(\pm 0.23)   $	& $(\pm 1.8)  $	 \\
 \hline
 \noalign{\smallskip}
 $m \sin i$		& [$M_\oplus$]		& $ 11.17$		& $ 47.9$		 \\
 			&				& $(\pm 0.66) $	& $(\pm 22.5) $	 \\
 $a$			& [AU]			& $ 0.0444 $		& $ 2.226 $		 \\
 			&				& $(\pm 0.0007)   $	& $(\pm 0.051)$	 \\
 \hline
 \noalign{\smallskip}
 $N_\mathrm{meas}$ &			& \multicolumn{2}{c}{100}  \\
 Span		& [days]			& \multicolumn{2}{c}{2615} \\
 rms			& [m\,s$^{-1}$]		& \multicolumn{2}{c}{1.39} \\
 $\chi_r^2$	&				& \multicolumn{2}{c}{2.65} \\
 \hline
 \end{tabular}
 \end{center}
 \end{table*}
 
 %----------------------------------------------------------------------
 
 %---------------------------  HD134606  ----------
 
 \begin{table*}
 \caption{Orbital and physical parameters of the planets orbiting HD\,134606 as obtained from a Keplerian fit to the data. }
 %\label{table_hd134606_par}
 \begin{center}
 \begin{tabular}{l l c c c}
 \hline \hline
 \noalign{\smallskip}
 {\bf Parameter}	& {\bf [unit]}		& {\bf HD 134606 b}	& {\bf HD 134606 c}	& {\bf HD 134606 d} \\
 \hline 
 \noalign{\smallskip}
 %Epoch		& [BJD]			& \multicolumn{3}{c}{2'454'783.40362208}    \\ 
 %$i$			& [deg]			& \multicolumn{3}{c}{$ 90 $ (fixed) }  \\  
 $V$			& [km\,s$^{-1}$]		& \multicolumn{3}{c}{$ 2.0510\,(\pm 0.0006) $}  \\
 \hline 
 \noalign{\smallskip}
 $P$			& [days]			& $12.083$		& $59.52$			& $459.3$		 \\ 
 			&				& $(\pm 0.010)$	& $(\pm 0.17)$		& $(\pm 8.3)$	 \\ 
 $e$			&				& $ 0.15 $			& $ 0.29 $			& $ 0.46 $		 \\ 
 			&				& $(\pm 0.10)$		& $ (\pm 0.20)$		& $(\pm 0.09)$	 \\ 
 $K$			& [m\,s$^{-1}$]		& $ 2.68 $		& $ 2.17 $		& $ 3.66 $		 \\  
 			&				& $(\pm 0.25)   $	& $(\pm 0.35)  $	& $(\pm 0.62)  $	 \\
 \hline
 \noalign{\smallskip}
 $m \sin i$		& [$M_\oplus$]		& $ 9.27$		& $ 12.14 $		& $ 38.52 $		 \\
 			&				& $(\pm 0.95) $	& $(\pm 1.67) $	& $(\pm 4.12)  $	 \\
 $a$			& [AU]			& $ 0.1023 $		& $ 0.2962 $		& $ 1.157 $		 \\
 			&				& $(\pm 0.0017)   $	& $(\pm 0.0049)$	& $(\pm 0.024)  $	 \\
 \hline
 \noalign{\smallskip}
 $N_\mathrm{meas}$ &			& \multicolumn{3}{c}{113}  \\
 Span		& [days]			& \multicolumn{3}{c}{2548} \\
 rms			& [m\,s$^{-1}$]		& \multicolumn{3}{c}{1.66} \\
 $\chi_r^2$	&				& \multicolumn{3}{c}{3.96} \\
 \hline
 \end{tabular}
 \end{center}
 \end{table*}
 
 %----------------------------------------------------------------------
 
 %---------------------------  HD136352  ----------
 
 \begin{table*}
 \caption{Orbital and physical parameters of the planets orbiting HD\,136352 as obtained from a Keplerian fit to the data. }
 %\label{table_hd136352_par}
 \begin{center}
 \begin{tabular}{l l c c c}
 \hline \hline
 \noalign{\smallskip}
 {\bf Parameter}	& {\bf [unit]}		& {\bf HD 136352 b}	& {\bf HD 136352 c}	& {\bf HD 136352 d} \\
 \hline 
 \noalign{\smallskip}
 %Epoch		& [BJD]			& \multicolumn{3}{c}{2'454'783.40362208}    \\ 
 %$i$			& [deg]			& \multicolumn{3}{c}{$ 90 $ (fixed) }  \\  
 $V$			& [km\,s$^{-1}$]		& \multicolumn{3}{c}{$ -68.7120\,(\pm 0.0005) $}  \\
 \hline 
 \noalign{\smallskip}
 $P$			& [days]			& $11.577$		& $27.582$		& $106.72$		 \\ 
 			&				& $(\pm 0.006)$	& $(\pm 0.023)$	& $(\pm 1.03)$	 \\ 
 $e$			&				& $ 0.18 $			& $ 0.16 $			& $ 0.43 $		 \\ 
 			&				& $(\pm 0.14)$		& $ (\pm 0.07)$		& $(\pm 0.24)$	 \\ 
 $K$			& [m\,s$^{-1}$]		& $ 1.77 $		& $ 2.82 $		& $ 1.68 $		 \\  
 			&				& $(\pm 0.22)   $	& $(\pm 0.23)  $	& $(\pm 0.47)  $	 \\
 \hline
 \noalign{\smallskip}
 $m \sin i$		& [$M_\oplus$]		& $ 5.28$			& $ 11.38 $		& $ 9.59 $		 \\
 			&				& $(\pm 0.62) $		& $(\pm 0.10) $		& $(\pm 1.86)  $	 \\
 $a$			& [AU]			& $ 0.0933 $		& $ 0.1665 $		& $ 0.411 $		 \\
 			&				& $(\pm 0.0015)   $	& $(\pm 0.0028)$	& $(\pm 0.007)  $	 \\
 \hline
 \noalign{\smallskip}
 $N_\mathrm{meas}$ &			& \multicolumn{3}{c}{129}  \\
 Span		& [days]			& \multicolumn{3}{c}{2601} \\
 rms			& [m\,s$^{-1}$]		& \multicolumn{3}{c}{1.35} \\
 $\chi_r^2$	&				& \multicolumn{3}{c}{2.72} \\
 \hline
 \end{tabular}
 \end{center}
 \end{table*}
 
 %----------------------------------------------------------------------
 
 %---------------------------  HD150433  ----------
 
 \begin{table*}
 \caption{Orbital and physical parameters of the planets orbiting HD\,150433 as obtained from a Keplerian fit to the data. }
 %\label{table_hd150433_par}
 \begin{center}
 \begin{tabular}{l l c }
 \hline \hline
 \noalign{\smallskip}
 {\bf Parameter}	& {\bf [unit]}		& {\bf HD 150433 b}	 \\
 \hline 
 \noalign{\smallskip}
 %Epoch		& [BJD]			& \multicolumn{1}{c}{2'454'783.40362208}    \\ 
 %$i$			& [deg]			& \multicolumn{1}{c}{$ 90 $ (fixed) }  \\  
 $V$			& [km\,s$^{-1}$]		& \multicolumn{1}{c}{$ -40.1141\,(\pm 0.0003) $}  \\
 \hline 
 \noalign{\smallskip}
 $P$			& [days]			& $1096$		 \\ 
 			&				& $(\pm 27)$	 \\ 
 $e$			&				& $ 0.0 $			\\ 
 			&				& $(fixed)$		 \\ 
 $K$			& [m\,s$^{-1}$]		& $ 3.85 $		 \\  
 			&				& $(\pm 0.42)   $	 \\
 \hline
 \noalign{\smallskip}
 $m \sin i$		& [$M_\oplus$]		& $ 53.5$		 \\
 			&				& $(\pm 6.3) $	 \\
 $a$			& [AU]			& $ 1.930 $		 \\
 			&				& $(\pm 0.045)   $	 \\
 \hline
 \noalign{\smallskip}
 $N_\mathrm{meas}$ &			& \multicolumn{1}{c}{51}  \\
 Span		& [days]			& \multicolumn{1}{c}{2129} \\
 rms			& [m\,s$^{-1}$]		& \multicolumn{1}{c}{1.73} \\
 $\chi_r^2$	&				& \multicolumn{1}{c}{3.75} \\
 \hline
 \end{tabular}
 \end{center}
 \end{table*}
 
 %----------------------------------------------------------------------
 
 %---------------------------  HD154088  ----------
 
 \begin{table*}
 \caption{Orbital and physical parameters of the planets orbiting HD\,154088 as obtained from a Keplerian fit to the data. }
 %\label{table_hd154088_par}
 \begin{center}
 \begin{tabular}{l l c }
 \hline \hline
 \noalign{\smallskip}
 {\bf Parameter}	& {\bf [unit]}		& {\bf HD 154088 b}	 \\
 \hline 
 \noalign{\smallskip}
 %Epoch		& [BJD]			& \multicolumn{1}{c}{2'454'783.40362208}    \\ 
 %$i$			& [deg]			& \multicolumn{1}{c}{$ 90 $ (fixed) }  \\  
 $V$			& [km\,s$^{-1}$]		& \multicolumn{1}{c}{$ 14.2972\,(\pm 0.0003) $}  \\
 \hline 
 \noalign{\smallskip}
 $P$			& [days]			& $18.596$		 \\ 
 			&				& $(\pm 0.021)$	 \\ 
 $e$			&				& $ 0.38 $			\\ 
 			&				& $(\pm 0.15)$		 \\ 
 $K$			& [m\,s$^{-1}$]		& $ 1.78 $		 \\  
 			&				& $(\pm 0.31)   $	 \\
 \hline
 \noalign{\smallskip}
 $m \sin i$		& [$M_\oplus$]		& $ 6.15$		 \\
 			&				& $(\pm 0.86) $	 \\
 $a$			& [AU]			& $ 0.1316 $		 \\
 			&				& $(\pm 0.0021)   $	 \\
 \hline
 \noalign{\smallskip}
 $N_\mathrm{meas}$ &			& \multicolumn{1}{c}{112}  \\
 Span		& [days]			& \multicolumn{1}{c}{1924} \\
 rms			& [m\,s$^{-1}$]		& \multicolumn{1}{c}{1.24} \\
 $\chi_r^2$	&				& \multicolumn{1}{c}{2.21} \\
 \hline
 \end{tabular}
 \end{center}
 \end{table*}
 
 %----------------------------------------------------------------------
 \clearpage
 %---------------------------  HD157172  ----------
 
 \begin{table*}
 \caption{Orbital and physical parameters of the planets orbiting HD\,157172 as obtained from a Keplerian fit to the data. }
 %\label{table_hd157172_par}
 \begin{center}
 \begin{tabular}{l l c }
 \hline \hline
 \noalign{\smallskip}
 {\bf Parameter}	& {\bf [unit]}		& {\bf HD 157172 b}	 \\
 \hline 
 \noalign{\smallskip}
 %Epoch		& [BJD]			& \multicolumn{1}{c}{2'454'783.40362208}    \\ 
 %$i$			& [deg]			& \multicolumn{1}{c}{$ 90 $ (fixed) }  \\  
 $V$			& [km\,s$^{-1}$]		& \multicolumn{1}{c}{$ -78.9210\,(\pm 0.0006) $}  \\
 \hline 
 \noalign{\smallskip}
 $P$			& [days]			& $104.84$		 \\ 
 			&				& $(\pm 0.13)$	 \\ 
 $e$			&				& $ 0.46 $			\\ 
 			&				& $(\pm 0.05)$		 \\ 
 $K$			& [m\,s$^{-1}$]		& $ 6.42 $		 \\  
 			&				& $(\pm 0.43)   $	 \\
 \hline
 \noalign{\smallskip}
 $m \sin i$		& [$M_\oplus$]		& $ 38.1$		 \\
 			&				& $(\pm 2.6) $	 \\
 $a$			& [AU]			& $ 0.416 $		 \\
 			&				& $(\pm 0.007)   $	 \\
 \hline
 \noalign{\smallskip}
 $N_\mathrm{meas}$ &			& \multicolumn{1}{c}{82}  \\
 Span		& [days]			& \multicolumn{1}{c}{2157} \\
 rms			& [m\,s$^{-1}$]		& \multicolumn{1}{c}{1.90} \\
 $\chi_r^2$	&				& \multicolumn{1}{c}{13.03} \\
 \hline
 \end{tabular}
 \end{center}
 \end{table*}
 
 %----------------------------------------------------------------------
 
 %---------------------------  HD189567  ----------
 
 \begin{table*}
 \caption{Orbital and physical parameters of the planets orbiting HD\,189567 as obtained from a Keplerian fit to the data. }
 %\label{table_hd189567_par}
 \begin{center}
 \begin{tabular}{l l c }
 \hline \hline
 \noalign{\smallskip}
 {\bf Parameter}	& {\bf [unit]}		& {\bf HD 189567 b}	 \\
 \hline 
 \noalign{\smallskip}
 %Epoch		& [BJD]			& \multicolumn{1}{c}{2'454'783.40362208}    \\ 
 %$i$			& [deg]			& \multicolumn{1}{c}{$ 90 $ (fixed) }  \\  
 $V$			& [km\,s$^{-1}$]		& \multicolumn{1}{c}{$ -10.4792\,(\pm 0.0004) $}  \\
 \hline 
 \noalign{\smallskip}
 $P$			& [days]			& $14.275$		 \\ 
 			&				& $(\pm 0.005)$	 \\ 
 $e$			&				& $ 0.23 $			\\ 
 			&				& $(\pm 0.14)$		 \\ 
 $K$			& [m\,s$^{-1}$]		& $ 3.02 $		 \\  
 			&				& $(\pm 0.33)   $	 \\
 \hline
 \noalign{\smallskip}
 $m \sin i$		& [$M_\oplus$]		& $ 10.03$		 \\
 			&				& $(\pm 1.04) $	 \\
 $a$			& [AU]			& $ 0.1099 $		 \\
 			&				& $(\pm 0.0018)   $	 \\
 \hline
 \noalign{\smallskip}
 $N_\mathrm{meas}$ &			& \multicolumn{1}{c}{166}  \\
 Span		& [days]			& \multicolumn{1}{c}{2818} \\
 rms			& [m\,s$^{-1}$]		& \multicolumn{1}{c}{2.64} \\
 $\chi_r^2$	&				& \multicolumn{1}{c}{9.09} \\
 \hline
 \end{tabular}
 \end{center}
 \end{table*}
 
 %----------------------------------------------------------------------
 
 %---------------------------  HD204313  ----------
 
 \begin{table*}
 \caption{Orbital and physical parameters of the planets orbiting HD\,204313 as obtained from a Keplerian fit to the data. }
 %\label{table_hd204313_par}
 \begin{center}
 \begin{tabular}{l l c }
 \hline \hline
 \noalign{\smallskip}
 {\bf Parameter}	& {\bf [unit]}		& {\bf HD 204313 c}	 \\
 \hline 
 \noalign{\smallskip}
 %Epoch		& [BJD]			& \multicolumn{1}{c}{2'454'783.40362208}    \\ 
 %$i$			& [deg]			& \multicolumn{1}{c}{$ 90 $ (fixed) }  \\  
 $V$			& [km\,s$^{-1}$]		& \multicolumn{1}{c}{$ -9.7419\,(\pm 0.0021) $}  \\
 \hline 
 \noalign{\smallskip}
 $P$			& [days]			& $34.873$		 \\ 
 			&				& $(\pm 0.039)$	 \\ 
 $e$			&				& $ 0.17 $			\\ 
 			&				& $(\pm 0.09)$		 \\ 
 $K$			& [m\,s$^{-1}$]		& $ 3.36 $		 \\  
 			&				& $(\pm 0.35)   $	 \\
 \hline
 \noalign{\smallskip}
 $m \sin i$		& [$M_\oplus$]		& $ 17.15$		 \\
 			&				& $(\pm 1.71) $	 \\
 $a$			& [AU]			& $ 0.2103 $		 \\
 			&				& $(\pm 0.0035)   $	 \\
 \hline
 \noalign{\smallskip}
 $N_\mathrm{meas}$ &			& \multicolumn{1}{c}{67}  \\
 Span		& [days]			& \multicolumn{1}{c}{1547} \\
 rms			& [m\,s$^{-1}$]		& \multicolumn{1}{c}{1.08} \\
 $\chi_r^2$	&				& \multicolumn{1}{c}{1.78} \\
 \hline
 \end{tabular}
 \end{center}
 \end{table*}
 
 %----------------------------------------------------------------------
 
 %---------------------------  HD215152  ----------
 
 \begin{table*}
 \caption{Orbital and physical parameters of the planets orbiting HD\,215152 as obtained from a Keplerian fit to the data. }
 %\label{table_hd215152_par}
 \begin{center}
 \begin{tabular}{l l c c }
 \hline \hline
 \noalign{\smallskip}
 {\bf Parameter}	& {\bf [unit]}		& {\bf HD 215152 b}	& {\bf HD 215152 c}	 \\
 \hline 
 \noalign{\smallskip}
 %Epoch		& [BJD]			& \multicolumn{2}{c}{2'454'783.40362208}    \\ 
 %$i$			& [deg]			& \multicolumn{2}{c}{$ 90 $ (fixed) }  \\  
 $V$			& [km\,s$^{-1}$]		& \multicolumn{2}{c}{$ -13.6810\,(\pm 0.0012) $}  \\
 \hline 
 \noalign{\smallskip}
 $P$			& [days]			& $7.283$		& $10.866$		 \\ 
 			&				& $(\pm 0.006)$	& $(\pm 0.014)$	 \\ 
 $e$			&				& $ 0.34 $			& $ 0.38 $			\\ 
 			&				& $(\pm 0.27)$		& $ (\pm 0.23)$		 \\ 
 $K$			& [m\,s$^{-1}$]		& $ 1.26 $		& $ 1.26 $		 \\  
 			&				& $(\pm 0.36)   $	& $(\pm 0.32)  $	 \\
 \hline
 \noalign{\smallskip}
 $m \sin i$		& [$M_\oplus$]		& $ 2.78$		& $ 3.10 $		 \\
 			&				& $(\pm 0.47) $	& $(\pm 0.48) $	 \\
 $a$			& [AU]			& $ 0.0652 $		& $ 0.0852 $		 \\
 			&				& $(\pm 0.0010)   $	& $(\pm 0.0014)$	 \\
 \hline
 \noalign{\smallskip}
 $N_\mathrm{meas}$ &			& \multicolumn{2}{c}{171}  \\
 Span		& [days]			& \multicolumn{2}{c}{2927} \\
 rms			& [m\,s$^{-1}$]		& \multicolumn{2}{c}{1.33} \\
 $\chi_r^2$	&				& \multicolumn{2}{c}{2.15} \\
 \hline
 \end{tabular}
 \end{center}
 \end{table*}
 
 %----------------------------------------------------------------------
 
 %---------------------------  HD215456  ----------
 
 \begin{table*}
 \caption{Orbital and physical parameters of the planets orbiting HD\,215456 as obtained from a Keplerian fit to the data. }
 %\label{table_hd215456_par}
 \begin{center}
 \begin{tabular}{l l c c }
 \hline \hline
 \noalign{\smallskip}
 {\bf Parameter}	& {\bf [unit]}		& {\bf HD 215456 b}	& {\bf HD 215456 c}	 \\
 \hline 
 \noalign{\smallskip}
 %Epoch		& [BJD]			& \multicolumn{2}{c}{2'454'783.40362208}    \\ 
 %$i$			& [deg]			& \multicolumn{2}{c}{$ 90 $ (fixed) }  \\  
 $V$			& [km\,s$^{-1}$]		& \multicolumn{2}{c}{$ -18.8541\,(\pm 0.0006) $}  \\
 \hline 
 \noalign{\smallskip}
 $P$			& [days]			& $191.99$		& $2277$		 \\ 
 			&				& $(\pm 0.73)$		& $(\pm 67)$	 \\ 
 $e$			&				& $ 0.15 $			& $ 0.19 $			\\ 
 			&				& $(\pm 0.10)$		& $ (\pm 0.11)$		 \\ 
 $K$			& [m\,s$^{-1}$]		& $ 3.62 $		& $ 3.89 $		 \\  
 			&				& $(\pm 0.31)   $	& $(\pm 0.44)  $	 \\
 \hline
 \noalign{\smallskip}
 $m \sin i$		& [$M_\oplus$]		& $ 32.21$		& $ 78.37 $		 \\
 			&				& $(\pm 2.92) $	& $(\pm 8.86) $	 \\
 $a$			& [AU]			& $ 0.652 $		& $ 3.394 $		 \\
 			&				& $(\pm 0.011)   $	& $(\pm 0.088)$	 \\
 \hline
 \noalign{\smallskip}
 $N_\mathrm{meas}$ &			& \multicolumn{2}{c}{96}  \\
 Span		& [days]			& \multicolumn{2}{c}{2615} \\
 rms			& [m\,s$^{-1}$]		& \multicolumn{2}{c}{1.79} \\
 $\chi_r^2$	&				& \multicolumn{2}{c}{5.41} \\
 \hline
 \end{tabular}
 \end{center}
 \end{table*}

  \end{document}